\documentclass[longauth]{aa}
\usepackage{graphicx}
\usepackage{natbib}
\usepackage{textcomp}
\usepackage{siunitx}
\usepackage{txfonts}
\usepackage[colorlinks=true]{hyperref}

\newcounter{RomC}

\def\mytarget#1{%
\vspace*{-3em}
\hypertarget{#1}{}
\vspace*{3em}
}

\input{macros.tex}

\begin{document} 

   \title{The Solar Orbiter Science Activity Plan}
   \subtitle{Translating solar and heliospheric physics questions into action}
    \author{I.~Zouganelis\inst{\ref{inst:esac}}\thanks{Corresponding author, e-mail: yannis.zouganelis@esa.int},
A.~De~Groof\inst{\ref{inst:esac}},
A.~P.~Walsh\inst{\ref{inst:esac}},
D.~R.~Williams\inst{\ref{inst:esac}},
D.~M\"uller\inst{\ref{inst:estec}},
O.~C.~St~Cyr\inst{\ref{inst:gsfc}},
F.~Auch\`ere\inst{\ref{inst:ias}},
D.~Berghmans\inst{\ref{inst:rob}},
A.~Fludra\inst{\ref{inst:ral}},
T.~S.~Horbury\inst{\ref{inst:imp}},
R.~A.~Howard\inst{\ref{inst:nrl}},
S.~Krucker\inst{\ref{inst:fhnws}},
M.~Maksimovic\inst{\ref{inst:lesia}},
C.~J.~Owen\inst{\ref{inst:mssl}},
J.~Rodr\'iguez-Pacheco\inst{\ref{inst:alc}},
M.~Romoli\inst{\ref{inst:unifi}},
S.~K.~Solanki\inst{\ref{inst:mps},\ref{inst:ssr}},
C.~Watson\inst{\ref{inst:esac}},
L.~Sanchez\inst{\ref{inst:esac}},
J.~Lefort\inst{\ref{inst:esac}},
P.~Osuna\inst{\ref{inst:esac}},
H.~R.~Gilbert\inst{\ref{inst:gsfc}},
T.~Nieves-Chinchilla\inst{\ref{inst:gsfc}},
L.~Abbo\inst{\ref{inst:inaf_to}},
O.~Alexandrova\inst{\ref{inst:lesia}},
A.~Anastasiadis\inst{\ref{inst:noa}},
V.~Andretta\inst{\ref{inst:inaf_np}},
E.~Antonucci\inst{\ref{inst:inaf_to}},
T.~Appourchaux\inst{\ref{inst:ias}},
A.~Aran\inst{\ref{inst:barc}},
C.~N.~Arge\inst{\ref{inst:gsfc}},
G.~Aulanier\inst{\ref{inst:lesia}},
D.~Baker \inst{\ref{inst:mssl}},
S.~D.~Bale\inst{\ref{inst:ssl}},
M.~Battaglia\inst{\ref{inst:fhnws}},
L.~Bellot~Rubio\inst{\ref{inst:iaa}},
A.~Bemporad\inst{\ref{inst:inaf_to}},
M.~Berthomier\inst{\ref{inst:lpp}},
K.~Bocchialini\inst{\ref{inst:ias}},
X.~Bonnin\inst{\ref{inst:lesia}},
A.~S.~Brun\inst{\ref{inst:cea}},
R.~Bruno\inst{\ref{inst:iaps}},
E.~Buchlin\inst{\ref{inst:ias}},
J.~B\"uchner\inst{\ref{inst:berlin}},
R.~Bucik\inst{\ref{inst:swrisa},\ref{inst:mps}},
F.~Carcaboso\inst{\ref{inst:alc}},
R.~Carr\inst{\ref{inst:esac}},
I.~Carrasco-Bl\'azquez\inst{\ref{inst:imp}},
B.~Cecconi\inst{\ref{inst:lesia}},
I.~Cernuda~Cangas\inst{\ref{inst:alc}},
C.~H.~K.~Chen\inst{\ref{inst:queenmary}},
L.~P.~Chitta\inst{\ref{inst:mps}},
T.~Chust\inst{\ref{inst:lpp}},
K.~Dalmasse\inst{\ref{inst:irap}},
R.~D'Amicis\inst{\ref{inst:iaps}},
V.~Da~Deppo\inst{\ref{inst:padova}},
R.~De~Marco\inst{\ref{inst:iaps}},
S.~Dolei\inst{\ref{inst:inaf_ca}},
L.~Dolla\inst{\ref{inst:rob}},
T.~Dudok~de~Wit\inst{\ref{inst:lpc2e}},
L.~van~Driel-Gesztelyi\inst{\ref{inst:mssl}},
J.~P.~Eastwood\inst{\ref{inst:imp}},
F.~Espinosa~Lara\inst{\ref{inst:alc}},
L.~Etesi\inst{\ref{inst:fhnws}},
A.~Fedorov\inst{\ref{inst:irap}},
F.~F\'elix-Redondo\inst{\ref{inst:esac}},
S.~Fineschi\inst{\ref{inst:inaf_to}},
B.~Fleck\inst{\ref{inst:fleck}},
D.~Fontaine\inst{\ref{inst:lpp}},
N.~J.~Fox\inst{\ref{inst:nasahq}},
A.~Gandorfer\inst{\ref{inst:mps}},
V.~G\'enot\inst{\ref{inst:irap}},
M.~K.~Georgoulis\inst{\ref{inst:acadath},\ref{inst:georgia}},
S.~Gissot\inst{\ref{inst:rob}},
A.~Giunta\inst{\ref{inst:ral}},
L.~Gizon\inst{\ref{inst:mps}},
R.~G\'omez-Herrero\inst{\ref{inst:alc}},
K.~Gontikakis\inst{\ref{inst:acadath}},
G.~Graham\inst{\ref{inst:esac}},
L.~Green\inst{\ref{inst:mssl}},
T.~Grundy\inst{\ref{inst:ral}},
M.~Haberreiter\inst{\ref{inst:pmod}},
L.~K.~Harra\inst{\ref{inst:pmod},\ref{inst:eth}},
D.~M.~Hassler\inst{\ref{inst:swri}},
J.~Hirzberger\inst{\ref{inst:mps}},
G.~C.~Ho\inst{\ref{inst:jhu}},
G.~Hurford\inst{\ref{inst:ssl}},
D.~Innes\inst{\ref{inst:mps}},
K.~Issautier\inst{\ref{inst:lesia}},
A.~W.~James\inst{\ref{inst:esac}},
N.~Janitzek\inst{\ref{inst:esac}},
M.~Janvier\inst{\ref{inst:ias}},
N.~Jeffrey\inst{\ref{inst:nuu}},
J.~Jenkins\inst{\ref{inst:mssl},\ref{inst:leuven}},
Y.~Khotyaintsev\inst{\ref{inst:uppsala}},
K.-L.~Klein,\inst{\ref{inst:lesia}},
E.~P.~Kontar\inst{\ref{inst:glasgow}},
I.~Kontogiannis\inst{\ref{inst:pots}},
C.~Krafft\inst{\ref{inst:lpp}},
V.~Krasnoselskikh\inst{\ref{inst:lpc2e}},
M.~Kretzschmar\inst{\ref{inst:lpc2e}},
N.~Labrosse\inst{\ref{inst:glasgow}},
A.~Lagg\inst{\ref{inst:mps}},
F.~Landini\inst{\ref{inst:inaf_to}},
B.~Lavraud\inst{\ref{inst:irap}},
I.~Leon\inst{\ref{inst:esac}},
S.~T.~Lepri\inst{\ref{inst:mich}},
G.~R.~Lewis\inst{\ref{inst:mssl}},
P.~Liewer\inst{\ref{inst:jpl}},
J.~Linker\inst{\ref{inst:predi}},
S.~Livi\inst{\ref{inst:swrisa}},
D.~M.~Long\inst{\ref{inst:mssl}},
P.~Louarn\inst{\ref{inst:irap}},
O.~Malandraki\inst{\ref{inst:noa}},
S.~Maloney\inst{\ref{inst:sp},\ref{inst:scp}},
V.~Martinez-Pillet\inst{\ref{inst:nso}},
M.~Martinovic\inst{\ref{inst:arizona},\ref{inst:lesia}},
A.~Masson\inst{\ref{inst:esac}},
S.~Matthews \inst{\ref{inst:mssl}},
L.~Matteini\inst{\ref{inst:imp}},
N.~Meyer-Vernet\inst{\ref{inst:lesia}},
K.~Moraitis\inst{\ref{inst:lesia},\ref{inst:ioa}},
R.~J.~Morton\inst{\ref{inst:nuu}},
S.~Musset\inst{\ref{inst:glasgow}},
G.~Nicolaou\inst{\ref{inst:mssl}},
A.~Nindos\inst{\ref{inst:ioa}},
H.~O'Brien\inst{\ref{inst:imp}},
D.~Orozco~Suarez\inst{\ref{inst:iaa}},
M.~Owens\inst{\ref{inst:reading}},
M.~Pancrazzi\inst{\ref{inst:inaf_fl}},
A.~Papaioannou\inst{\ref{inst:noa}},
S.~Parenti\inst{\ref{inst:ias}},
E.~Pariat\inst{\ref{inst:lesia}},
S.~Patsourakos\inst{\ref{inst:ioa}},
D.~Perrone\inst{\ref{inst:asi}},
H.~Peter\inst{\ref{inst:mps}},
R.~F.~Pinto\inst{\ref{inst:irap},\ref{inst:cea}},
C.~Plainaki\inst{\ref{inst:asi}},
D.~Plettemeier\inst{\ref{inst:dresden}},
S.~P.~Plunkett\inst{\ref{inst:nasahq}},
J.~M.~Raines\inst{\ref{inst:mich}},
N.~Raouafi\inst{\ref{inst:jhu}},
H.~Reid\inst{\ref{inst:mssl}},
A.~Retino\inst{\ref{inst:lpp}},
L.~Rezeau\inst{\ref{inst:lpp}},
P.~Rochus\inst{\ref{inst:csl}},
L.~Rodriguez\inst{\ref{inst:rob}},
L.~Rodriguez-Garcia\inst{\ref{inst:alc}},
M.~Roth\inst{\ref{inst:kis}},
A.~P.~Rouillard\inst{\ref{inst:irap}},
F.~Sahraoui\inst{\ref{inst:lpp}},
C.~Sasso\inst{\ref{inst:inaf_np}},
J.~Schou\inst{\ref{inst:mps}},
U.~Sch\"uhle\inst{\ref{inst:mps}},
L.~Sorriso-Valvo\inst{\ref{inst:bari},\ref{inst:quito}},
J.~Soucek\inst{\ref{inst:iapprague}},
D.~Spadaro\inst{\ref{inst:inaf_ca}},
M.~Stangalini\inst{\ref{inst:asi}},
D.~Stansby\inst{\ref{inst:mssl}},
M.~Steller\inst{\ref{inst:graz}},
A.~Strugarek\inst{\ref{inst:cea}},
\v{S}. \v{S}tver\'ak\inst{\ref{inst:iapprague},\ref{inst:aiprague}},
R.~Susino\inst{\ref{inst:inaf_to}},
D.~Telloni\inst{\ref{inst:inaf_to}},
C.~Terasa\inst{\ref{inst:cau}},
L.~Teriaca\inst{\ref{inst:mps}},
S.~Toledo-Redondo\inst{\ref{inst:murcia}},
J.~C.~del~Toro~Iniesta\inst{\ref{inst:iaa}},
G.~Tsiropoula\inst{\ref{inst:noa}},
A.~Tsounis \inst{\ref{inst:esac}},
K.~Tziotziou\inst{\ref{inst:noa}},
F.~Valentini\inst{\ref{inst:calabria}},
A.~Vaivads\inst{\ref{inst:stockholm}},
A.~Vecchio\inst{\ref{inst:lesia},\ref{inst:radionl}},
M.~Velli\inst{\ref{inst:jpl}},
C.~Verbeeck\inst{\ref{inst:rob}},
A.~Verdini\inst{\ref{inst:unifi}},
D.~Verscharen\inst{\ref{inst:mssl},\ref{inst:hamp}},
N.~Vilmer\inst{\ref{inst:lesia}},
A.~Vourlidas\inst{\ref{inst:jhu}},
R.~Wicks\inst{\ref{inst:nuu}},
R.~F.~Wimmer-Schweingruber\inst{\ref{inst:cau}},
T.~Wiegelmann\inst{\ref{inst:mps}},
P.~R.~Young \inst{\ref{inst:gsfc}, \ref{inst:nuu}},
A.~N.~Zhukov\inst{\ref{inst:rob},\ref{inst:sinp}}}

\authorrunning{Zouganelis et al.}
   \institute{European Space Agency, ESAC, Camino Bajo del Castillo s/n, Urb. Villafranca del Castillo, 28692 Villanueva de la Ca\~nada, Madrid, Spain\label{inst:esac} 
\and
European Space Agency, ESTEC, P.O. Box 299, 2200 AG Noordwijk, The Netherlands\label{inst:estec}
\and 
Solar Physics Laboratory, Heliophysics Science Division, NASA Goddard Space Flight Center, Greenbelt, MD, 2077\label{inst:gsfc}
\and
Universit\'e Paris-Saclay, CNRS, Institut d'Astrophysique Spatiale, 91405 Orsay, France\label{inst:ias}
\and
Solar-Terrestrial Centre of Excellence -- SIDC, Royal Observatory of Belgium, Ringlaan -3- Av. Circulaire, 1180 Brussels, Belgium\label{inst:rob}
\and
RAL Space, STFC Rutherford Appleton Laboratory, Harwell, Didcot, OX11 0QX, UK\label{inst:ral}
\and
Department of Physics, Imperial College London, London SW7 2AZ, UK\label{inst:imp}
\and
Naval Research Laboratory, Space Science Division, Washington, DC 20375, USA\label{inst:nrl}
\and
University of Applied Sciences and Arts Northwestern Switzerland, CH-5210 Windisch, Switzerland\label{inst:fhnws}
\and
LESIA, Observatoire de Paris, Universit\'e PSL, CNRS, Sorbonne Universit\'e, Universit\'e de Paris, 5 place Jules Janssen, 92195 Meudon, France\label{inst:lesia}\newpage
\and
Mullard Space Science Laboratory, University College London, Holmbury St Mary, Dorking, RH5 6NT, United Kingdom\label{inst:mssl}
\and
Universidad de Alcal\'a, Space Research Group, 28805 Alcal\'a de Henares, Spain\label{inst:alc}
\and
Dipartimento di Fisica e Astronomia, Universit\`a degli Studi di Firenze, Largo E. Fermi 2, 50125, Firenze, Italy\label{inst:unifi}
\and
Max-Planck-Institut f\"ur Sonnensystemforschung, Justus-von-Liebig-Weg 3, 37077 G\"ottingen, Germany\label{inst:mps}
\and
School of Space Research, Kyung Hee University, Yongin, Gyeonggi-Do, 446-701, Republic of Korea\label{inst:ssr}
\and
INAF Osservatorio Astrofisico di Torino, Via Osservatorio 20, 10025, Pino Torinese, Italy\label{inst:inaf_to}
\and
Institute for Astronomy, Astrophysics, Space Applications and Remote Sensing (IAASARS), National Observatory of Athens, Metaxa and Vas. Pavlou St., Penteli, 15236 Athens, Greece\label{inst:noa}
\and
INAF Osservatorio Astronomico di Capodimonte, Salita Moiariello 16, 80131, Napoli, Italy\label{inst:inaf_np}
\and
Departament de F\'isica Qu\`antica i Astrof\'isica, Institut de Ci\`encies del Cosmos (ICCUB), Universitat de Barcelona (IEEC-UB),Spain\label{inst:barc}
\and
Space Sciences Laboratory and Physics Department, University of California, Berkeley, USA\label{inst:ssl}
\and
Instituto de Astrofísica de Andalucía (IAA-CSIC), Apdo. de Correos 3004, E-18080 Granada, Spain\label{inst:iaa}
\and
LPP, CNRS, Ecole Polytechnique, Sorbonne Universit\'e, Observatoire de Paris, Universit\'e Paris-Saclay, PSL Research University, Paris, France\label{inst:lpp}
\and
AIM, CEA, CNRS, Universit\'e Paris-Saclay, Universit\'e de Paris, Sorbonne Paris Cit\'e, F-91191 Gif-sur-Yvette, France\label{inst:cea}
\and
INAF Istituto di Astrofisica e Planetologia Spaziale, Via Fosso del Cavaliere 100, 00133 Roma, Italy\label{inst:iaps}
\and
Center for Astronomy and Astrophysics, Berlin Institute of Technology (Technische Universit\"at Berlin), 10623 Berlin, Germany\label{inst:berlin}
\and
Southwest Research Institute, 6220 Culebra Rd, San Antonio, TX 78238, USA\label{inst:swrisa}
\and
School of Physics and Astronomy, Queen Mary University of London, London E1 4NS, UK\label{inst:queenmary}
\and
IRAP, Universit\'e de Toulouse, CNRS, CNES, UPS, F-31028 Toulouse, France\label{inst:irap}
\and
CNR-IFN LUXOR, Via Trasea 7, 35131, Padova, Italy\label{inst:padova}
\and
INAF Osservatorio Astrofisico di Catania, Via S. Sofia 78, 95123, Catania, Italy\label{inst:inaf_ca}
\and
LPC2E, UMR7328 CNRS and University of Orl\'eans, 3a av. de la recherche scientifique, Orl\'eans, France\label{inst:lpc2e}
\and
ESA Science and Operations Department, c/o NASA/GSFC Code 671, Greenbelt, MD 20771, USA\label{inst:fleck}
\and
NASA Headquarters, Washington, DC20546, USA\label{inst:nasahq}
\and
Research Center for Astronomy and Applied Mathematics of the Academy of Athens, 11527 Athens, Greece\label{inst:acadath}
\and
Department of Physics and Astronomy, Georgia State University, Atlanta, GA 30303, USA\label{inst:georgia}
\and
Physikalisch-Meteorologisches Observatorium Davos, World Radiation Center, 7260, Davos Dorf\label{inst:pmod}
\and
ETH-Zürich, IPA, Hönggerberg campus, Zürich, Switzerland. \label{inst:eth}
\and
Southwest Research Institute, 1050 Walnut Street, Boulder, CO, USA\label{inst:swri}
\and
The Johns Hopkins University Applied Physics Laboratory, Laurel, MD, USA\label{inst:jhu}
\and
Department of Mathematics, Physics and Electrical Engineering, Northumbria University, Newcastle upon Tyne, NE1 8ST, UK\label{inst:nuu}
\and
Centre for mathematical Plasma-Astrophysics, Celestijnenlaan 200B, 3001 Leuven, KU Leuven, Belgium\label{inst:leuven}
\and
Swedish Institute of Space Physics, Uppsala, Sweden\label{inst:uppsala}
\and
SUPA, School of Physics and Astronomy, University of Glasgow, Glasgow G12 8QQ, UK\label{inst:glasgow}
\and
Leibniz-Institut f\"{u}r Astrophysik Potsdam (AIP), An der Sternwarte 16, 14482 Potsdam, Germany\label{inst:pots}
\and
Department of Climate and Space Sciences and Engineering, University of Michigan, Ann Arbor, Michigan, USA\label{inst:mich}
\and
Jet Propulsion Laboratory, California Institute of Technology, Pasadena, CA 91109, USA\label{inst:jpl}\newpage
\and
Predictive Science Inc., 9990 Mesa Rim Road, Suite 170, San Diego, CA, USA 92121\label{inst:predi}
\and
School of Physics, Trinity College Dublin, Dublin 2, Ireland\label{inst:sp}
\and
School of Cosmic Physics, Dublin Institute for Advanced Studies, Dublin D02 XF85, Ireland\label{inst:scp}
\and
National Solar Observatory, 3665 Discovery Drive, Boulder, CO 80303 United States\label{inst:nso}
\and
Lunar and Planetary Laboratory, University of Arizona, Tucson, AZ 85721, USA\label{inst:arizona}
\and
Physics Department, University of Ioannina, Ioannina GR-45110, Greece\label{inst:ioa}
\and
Department of Meteorology, University of Reading, RG6 6BB, UK\label{inst:reading}
\and
INAF Osservatorio Astrofisico di Arcetri, Largo E. Fermi 5, 50125, Firenze, Italy\label{inst:inaf_fl}
\and
ASI - Italian Space Agency, via del Politecnico snc, 00133 Rome, Italy\label{inst:asi}
\and
Chair for RF-Engineering, Technische Universit\"{a}t Dresden, 01069 Dresden, Germany\label{inst:dresden}
\and
Centre Spatial de Liège, Université de Liège, Av. du Pré-Aily, 4031 Angleur, Belgium\label{inst:csl}
\and
Leibniz-Institut f\"{u}r Sonnenphysik, Sch\"{o}neckstr. 6, 79104 Freiburg, Germany\label{inst:kis}
\and
CNR - Istituto per la Scienza e Tecnologia dei Plasmi, Via Amendola 122/D, 70126 Bari, Italy\label{inst:bari}\newpage
\and
Departamento de F\'isica, Escuela Polit\'ecnica Nacional, Ladr\'on de Guevara 253, 170517 Quito, Ecuador\label{inst:quito}
\and
Institute of Atmospheric Physics, Czech Academy of Sciences, Prague, Czechia\label{inst:iapprague}
\and
Space Research Institute, Austrian Academy of Sciences, Graz, Austria\label{inst:graz}
\and
Astronomical Institute, Czech Academy of Sciences, Prague, Czechia\label{inst:aiprague}
\and
Division for Extraterrestrial Physics, Institute for Experimental and Applied Physics (IEAP), Christian Albrechts University at Kiel, Leibnizstr. 11, 24118 Kiel, Germany\label{inst:cau}
\and
Department of Electromagnetism and Electronics, University of Murcia, Murcia, Spain\label{inst:murcia}
\and
Dipartimento di Fisica, Universit\`a della Calabria, 87036 Rende (CS), Italy\label{inst:calabria}
\and
Space and Plasma Physics, School of Electrical Engineering and Computer Science, KTH Royal Institute of Technology, 100 44 Stockholm, Sweden\label{inst:stockholm}
\and
Radboud Radio Lab, Department of Astrophysics, IMAPP-Radboud University, 6500GL Nijmegen, The Netherlands\label{inst:radionl}
\and
Space Science Center, University of New Hampshire, 8 College Road, Durham NH 03824, United States\label{inst:hamp}
\and
Skobeltsyn Institute of Nuclear Physics, Moscow State University, Moscow, Russia\label{inst:sinp}
}

   \date{Received xxx / Accepted xxx}
  
   \abstract{Solar Orbiter is the first space mission observing the solar plasma both in situ and remotely, from a close distance, in and out of the ecliptic. The ultimate goal is to understand how the Sun produces and controls the heliosphere, filling the Solar System and driving the planetary environments. With six remote-sensing and four in-situ instrument suites, the coordination and planning of the operations are essential to address the following four top-level science questions: (1) What drives the solar wind and where does the coronal magnetic field originate?; (2) How do solar transients drive heliospheric variability?; (3) How do solar eruptions produce energetic particle radiation that fills the heliosphere?; (4) How does the solar dynamo work and drive connections between the Sun and the heliosphere? Maximising the mission's science return requires considering the characteristics of each orbit, including the relative position of the spacecraft to Earth (affecting downlink rates), trajectory events (such as gravitational assist manoeuvres), and the phase of the solar activity cycle. Furthermore, since each orbit's science telemetry will be downloaded over the course of the following orbit, science operations must be planned at mission level, rather than at the level of individual orbits. It is important to explore the way in which those science questions are translated into an actual plan of observations that fits into the mission, thus ensuring that no opportunities are missed. First, the overarching goals are broken down into specific, answerable questions along with the required observations and the so-called Science Activity Plan (SAP) is developed to achieve this. The SAP groups objectives that require similar observations into Solar Orbiter Observing Plans (SOOPs), resulting in a strategic, top-level view of the optimal opportunities for science observations during the mission lifetime. This allows for all four mission goals to be addressed. In this paper, we introduce Solar Orbiter's SAP through a series of examples and the strategy being followed.
}%{}{}{}{}
 
 % \abstract
  % [A&A: 300 words max.!]
%   {Context.
%}
%{}
 %    {}
    
\keywords{Sun: general -- Sun: magnetic fields -- Sun: activity -- Sun: atmosphere -- Sun: solar wind -- Methods: observational}

\maketitle

\section{Introduction}
\label{sect-intro}

Coordination and planning will be the key to the scientific success of Solar Orbiter, the new solar and heliospheric mission of the European Space Agency (ESA) with strong NASA participation, which was launched in February 2020 \citep{Mueller2020a}. This need for planning originates from the science goals of the mission, the capabilities of the platform and the payload, together with the telemetry constraints due to the deep-space characteristics of the orbit, and this has driven the design of the operations concept. To maximise the science return, observations must be coordinated between the remote-sensing and in-situ payload, amongst the remote-sensing instruments themselves \citep{Auchere2020a}, and also amongst the in-situ instruments \citep{Walsh2020}. The science return of Solar Orbiter and other contemporary space missions, particularly the Parker Solar Probe \citep{fox_solar_2016} and Bepi Colombo \citep{2010P&SS...58....2B}, and ground-based facilities such as the Daniel K. Inouye Solar Telescope DKIST (e.g. \citealt{2016AN....337.1064T}) or the Square Kilometer Array \citep{2019AdSpR..63.1404N} will also be enhanced through cross-facility coordination, within the constraints under which each facility operates \citep{Velli2020a}.

The Solar Orbiter Science Activity Plan (SAP) has been built with contributions from a significant portion of the international solar physics and heliophysics community. In a structured way, it describes the activities that are to be carried out by the instruments throughout all science mission phases in order to fulfil the science goals of the mission. It tracks how overarching science goals are mapped to more specific objectives, which are in turn mapped to observational activities scheduled at specific times during the mission. As context information, it also contains instruments' operation scenarios and resource modelling, along with a description of all mission phases. One of the main reasons for mission level planning is to ensure that there will be enough opportunities to address all of the mission's science objectives in an optimal way and that transferring an optimal amount of data to ground will be possible. More importantly, we need to make sure that if, for example, a unique opportunity exists for a specific science goal at a given date or configuration, this goal will be given higher priority. From this point of view, priorities have to not only be given based on the perceived importance of a science objective, which may change as science evolves, but also on other circumstances detailed in this paper.

In the following sections, we explain the observing campaigns, called SOOPs for `Solar Orbiter Observing Plans', which are the building blocks of the SAP,  and the strategy that we will adopt in building the plan during the next decade within the Science Working Team. The purpose of this paper is to provide the scientific community with a reference guide for the different SOOPs, especially for those scientists who are not directly involved in the science planning, but will be the ultimate users of the data. 

\section{Solar Orbiter science objectives}
\label{sect-objectives}

Over the past 30 years, an international effort to understand the Sun and heliosphere has been undertaken with an array of spacecraft carrying out both remote observations at visible, UV, X and Gamma-ray wavelengths, as well as in-situ observations of interplanetary plasmas, particles, and fields. Combined and coordinated observations from missions such as Helios \citep{1977JGZG...42..551P}, Ulysses \citep{1992A&AS...92..207W}, Yohkoh \citep{1992Sci...258..618A}, SOHO \citep{1995SoPh..162....1D}, TRACE \citep{1994SSRv...70..119S}, RHESSI \citep{2002SoPh..210....3L}, Hinode \citep{2007SoPh..243....3K}, STEREO \citep{2008SSRv..136....5K}, SDO \citep{2012SoPh..275....3P} and IRIS \citep{2014SoPh..289.2733D} as well as from ground-based observatories such as Gregor \citep{2012AN....333..796S}, GST \citep{2010AN....331..636C}, NVST \citep{2014RAA....14..705L}, SST \citep{2003SPIE.4853..341S} and ALMA \citep{2017SoPh..292...87S,2017SoPh..292...88W} have resulted in a leap forward in our understanding of the Sun and heliosphere, and have proven that critical progress in understanding the physics requires a synergy of remote and in-situ observations.

Solar Orbiter builds on previous missions and lessons learned to be able to address the central question of heliophysics: `how does the Sun create and control the heliosphere and why does solar activity change with time?' This, in turn, is a fundamental part of the second science question of ESA's Cosmic Vision programme: `How does the solar system work?'. Solar Orbiter is specifically designed to identify the origins and causes of the solar wind, the heliospheric magnetic field, solar energetic particles, transient interplanetary disturbances, and the Sun's magnetic field itself.

The supersonic solar wind, driven by dynamic plasma and magnetic processes near the Sun's surface, expands to surround the solar system's planets and the space far beyond. In the solar interior, the solar dynamo drives magnetic fields whose buoyancy brings them to the surface where they form huge arcades of loops, which store enormous amounts of magnetic energy. These magnetic loops are stretched and sheared by the Sun's differential rotation and transient surface processes, sometimes erupting in bright explosions, which eject magnetic structures and accelerate energetic particles that propagate into the solar system, occasionally impacting Earth and its magnetic shield with disruptive effects on space and terrestrial systems. Understanding the complex physical processes at work in this system is the central goal of heliophysics. Since the Sun and the heliosphere are typical of many late-type main-sequence stars and their astrospheres, these studies are relevant to astrophysics, but are unique since the Sun alone is close enough for detailed study.

Although Earth's vantage point at 1\,au is close by astrophysical measures, it has long been known that much of the crucial physics in the formation and activity of the heliosphere takes place much closer to the Sun, and that by the time magnetic structures, shocks, energetic particles and the solar wind pass by Earth, they have already evolved and in many cases mixed, blurring the signatures of their origin. With the proven effectiveness of combined remote and in-situ studies, critical new advances are expected to be achieved by Solar Orbiter, which takes this principle much closer to the Sun. With Solar Orbiter's unique combination of close distance to the Sun (minimum perihelion of 0.28\,au), and out-of-ecliptic vantage points (reaching above 17$^\circ$  heliographic latitude during its nominal mission phase and above 30$^\circ$ during the extended mission phase), solar sources will be identified and studied accurately and combined with in-situ observations of solar wind, shocks, energetic particles, before they have a chance to evolve significantly.

This big challenging question has been expressed in the form of the following four science questions \citep{Mueller2020a}:

	(1) What drives the solar wind and where does the coronal magnetic field originate?
	
	(2) How do solar transients drive heliospheric variability?
	
	(3) How do solar eruptions produce energetic particle radiation that fills the heliosphere?
	
	(4) How does the solar dynamo work and drive connections between the Sun and the heliosphere?

The Solar Orbiter science community has defined the detailed science activities that are required to bring closure to these four mission science questions. Around 500 objectives have been defined that cover all the above questions. These are described in detail in the Science Activity Plan pages of the Solar Orbiter science website\footnote{\url{https://www.cosmos.esa.int/web/solar-orbiter}} and will be continuously updated as new knowledge becomes available.

\section{Solar Orbiter mission planning overview}

The Solar Orbiter mission is unique from the operations point of view (see \citealt{Sanchez2020}), in that it will be the first solar mission to carry Sun-observing telescopes significantly closer to the Sun than previous missions, as close as 0.28\,au. In that sense, the operations concept is very different from those of previous missions such as SOHO, STEREO, Hinode, or SDO. As for all encounter and interplanetary missions, the various spacecraft resources are limited, a complexity that has to be taken into account in the operations philosophy. While the general mission planning approach for all routine science operations of Solar Orbiter builds on the experience of ESA's precursor solar system missions Mars Express, Venus Express, and Rosetta, a difference with respect to planetary missions is the highly dynamic nature of the Sun. In addition, while the in-situ instruments will operate continuously, this is not true for the remote-sensing instruments. During each orbit, the complete instrument suite will be operated during three 10-day windows, generally centred around closest approach, and at the minimum and maximum heliographic latitudes (see Fig. 1 for an example). These windows are called Remote-Sensing Windows (RSWs) and are central to the science planning. The resulting requirements on the science operations planning for remote-sensing observations are summarised below together with the mission planning cycle. Given the short time scales on which the targets of remote-sensing observations (such as solar active regions) change, together with the narrow fields of view of the high-resolution imaging telescopes (which cover less than 3\% of the solar disc at perihelion), turn-around times of, at most, three days between defining the pointing and executing the observations are required during RSWs.

Following a short phase of spacecraft and instrument commissioning, the Cruise Phase (CP) starts in mid-June 2020, which coincides with the first perihelion at a distance of 0.51\,au. During this phase, only the in-situ instruments will be operating nominally. The remote-sensing instruments will be occasionally switched on for calibration activities to ensure their readiness for the Nominal Mission Phase (NMP) on 26 November 2021, which marks the start of the nominal science operations of the entire payload. The NMP will last for five years until the 24 December 2026 after which an Extended Mission Phase (EMP) would be planned for another three years. The high latitudes will be progressively reached, exceeding 18$^\circ$ on 22 March 2025, 25$^\circ$ on 28 January 2027 and 33$^\circ$ on 23 July 2029. The fact that high latitudes are only reached towards the late phases has a direct impact on the planning of science objectives as this will be discussed in the following sections.

%\subsection{Mission planning cycle}

For each mission phase (CP, NMP, EMP), a baseline science plan is established and documented in the SAP before each mission phase commences. This plan takes into account the general characteristics and major constraints of each orbit. The SAP is a \textit{living} document in the sense that it will be frequently updated as science activities are worked out in detail and feedback from earlier observations (and their planning) is utilised for the planning of later orbits. It is important to reiterate that top-level science operations planning for the mission needs to be done well in advance due to the continuously changing orbital configuration, and to the fundamental constraints on Solar Orbiter's data downlink volume and onboard storage in the Solid State Mass Memory (SSMM). In fact, because of its unique orbit in the inner heliosphere, Solar Orbiter is a fundamentally different mission compared to solar missions such as SOHO or SDO that are near-Earth or the STEREO spacecraft that are far from Earth but still at 1\,au from the Sun. It should be compared to a planetary or deep-space mission, which entails various operational constraints. 

Because of the varying distance, $r$, between Earth and the spacecraft, which can be up to 2\,au (when the spacecraft is at 1\,au from the Sun, but in opposition to Earth), the telemetry will be highly variable, with a 1/$r^2$ dependence. Due to limited onboard data storage, there will be times during some orbits when the SSMM will be full and any new observations would overwrite older data before the latter can be downlinked. To prevent this, it is important to plan the observations accordingly, which is one objective of the SAP, but also to implement alternative solutions such as prioritisation of data downlink, onboard data processing and compression whenever possible. We refer the reader to the individual instrument papers (this issue\footnote{\url{https://www.aanda.org/component/toc/?task=topic&id=1082}}) for details on these solutions that are implemented for some of the instruments to variable extents.
At the mission level, the above constraints require feasibility studies of planned operations taking into account the expected time-dependent telemetry downlink profile as well as SSMM load levels. The mission planning cycle for the routine science operations phase is therefore divided into the following different levels.

\subsection{Long-Term Planning (LTP)}

Long-Term Planning (LTP) covers six-month periods starting in either January and July; this is driven by the schedule of ground station allocations. In this LTP, the ten instruments build a feasible, coordinated plan that will address the science objectives decided for that six month period in the SAP. In the early part of the mission, six months approximately correspond to an orbit, however later on, orbits will be shorter, so this correspondence will not always hold. In that case, the LTPs will still cover a period of six months, without any impact on the overall planning or operations.

\subsection{Short-Term Planning (STP)}

Short-Term Planning (STP) will result in the generation of detailed schedules of commands for the spacecraft and for the ground stations. This process will take place typically once a week covering one-week timeframe. At STP level, the instrument activities can be modified, provided they fit into the resource envelope defined at LTP level.

\subsection{Very-Short-Term Planning (VSTP)}

In the case of RSWs, VSTP may be required to add flexibility to the pointing of the spacecraft and of all remote-sensing instruments, which do not have independent pointing capabilities, (see \citealt{Auchere2020a}). At this phase, even if all commands have been stored onboard of the spacecraft, the mission maintains flexibility to respond to the dynamic nature of solar activity by pointing the high-resolution cameras, which are co-aligned with each other, at the locations of the Sun with the highest science priority for that time. This planning level, with turn-around times of three days or less between observations and execution of the new Pointing Request (PTR), is required for RSWs in which features on the solar disc, such as active regions, shall be tracked over time. These turn-around times are mainly due to the short lifetimes and non-deterministic evolution of targets on the Sun. The absolute pointing error (APE) of the spacecraft, which depends on the spacecraft's temperature and can be relatively large compared to the fields of view of the high-resolution imaging telescopes, needs also to be taken into account for every new PTR. The VSTP consists of (i) initial target selection and (ii) updates to the pointing.

Prior to the start of a RSW, a limited set of precursor observations with the full-disc imaging telescopes of the EUI, SO/PHI and Metis instruments is performed and downlinked with high priority. Based on the downlinked data, the target for the start of the RSW will be defined. This step is required to make a decision on the pointing of the spacecraft and, in turn, of the high-resolution imaging telescopes. In case the orbital configuration permits making this decision by means of other observations (e.g. using ground-based telescopes), this step can be omitted.

During the course of a RSW, a limited set of daily low-latency data, consisting of full-disc images, will be downlinked with high priority. Based on the evaluation of these images, the pointing may be updated by uploading a PTR.

\section{Solar Orbiter Observing Plans (SOOPs)}

Solar Orbiter has been designed to achieve its science goals through coordinate observations between in-situ and remote-sensing instruments and SOOPs are the necessary tools to ensure effective coordination. Development of SOOPs took advantage of the experience with similar tools developed for previous missions such as the SOHO JOPs (Joint Observing Programmes) and the Hinode HOPs (Hinode Operation Plans). 

Similar to Solar Orbiter, SOHO has an integrated package of complimentary instruments that were operated in a coordinated manner and the acquired data analysed in a correlative and cooperative spirit. To coordinate the various, independently operated and independently pointing instruments, the SOHO team created the system of SOHO Joint Observing Programmes (JOPs)\footnote{\url{https://soho.nascom.nasa.gov/soc/JOPs}}, which grew to often include resources and assets from beyond the mission itself, such as other space missions, sounding rockets and ground-based facilities. Over 200 JOPs were proposed by the community and executed (many of them multiple times).

Hinode's mission operations concept is much more tightly coordinated than SOHO's, because all fields of view are co-pointed. Its analogue of the JOP, the Hinode Operation Plan (HOP)\footnote{\url{https://hinode.msfc.nasa.gov/hops.html}} has been implemented with the following three objectives in mind: to establish importance of observations to the mission, by scheduling the observations in advance; to ensure the timely coordination of the three instruments (and, in the case of its full-disc X-ray telescope, to use the correct sub-field of the Field of View); and, most commonly, to coordinate with external facilities. External coordination has been performed since the very first HOP. After the launch of IRIS in 2013, the HOP system was extended to include IRIS as an integral part of the planning system and is now called `IRIS and Hinode Operation Plan' (IHOP).

JOPs and (I)HOPs were reused multiple times by their respective missions, as the rules of communication and coordination became more familiar, after being established in the first attempts. By analogy to JOPs and (I)HOPs, the Solar Orbiter SOOP is a set of common operations by multiple instruments, that is, a collection of instrument modes and parameters designed to address a specific science objective, and a SOOP can be reused during different orbital opportunities. However, a specific SOOP can also be used to address other science goals than the ones for which it has been primarily designed. For instance, a SOOP designed to study the dynamics of an active region with two different imagers at high spatial resolution and high temporal cadence can be used for sub-objectives as diverse as the study of the structure of a coronal mass ejection or the understanding of the energy release during a flare. This enables the definition of 36 SOOPs to address more than 500 sub-objectives. This simplifies greatly the science operations and makes them more effective. Most SOOPs involve multiple instruments, since coordinated campaigns are needed for most of the objectives. However, not every instrument is necessarily required in each SOOP: some instruments can make individual observations when they are not needed for the coordinated campaigns; likewise, multiple SOOPs can be executed in parallel if technically possible. 

In this section we describe all the SOOPs defined prior to the launch of the mission. We note that this is an ever-evolving list as more possibilities may appear during the mission as our understanding of the instruments' performance enhances and our scientific knowledge evolves. The current list together with detailed descriptions will always be available on the Solar Orbiter science website\footnote{\url{https://www.cosmos.esa.int/web/solar-orbiter}}. For the names of the SOOPs we use the following format: \emph{A\_B\_res\_cad\_description}. \emph{A} can take the values \emph{IS} for in-situ only observations, that is, only the in-situ instruments participate in this SOOP, \emph{R} for remote-sensing only observations, and \emph{L} for SOOPs that are designed for linking both in-situ and remote-sensing observations. $B$ can take the values \emph{FULL} for imaging of the entire solar disc, \emph{SMALL} for a small field of view (e.g. for imaging a specific active region) and \emph{BOTH} when the above are combined. \emph{res} refers to the spatial resolution of the imagers and it can take the values \emph{LRES}, \emph{MRES} and \emph{HRES} for low, medium and high spatial resolution respectively. Same for \emph{cad} but for the temporal cadence which can take the values \emph{LCAD}, \emph{MCAD} and \emph{HCAD} for low, medium and high cadence respectively. Lastly, a \emph{description} is added with plain language words (such as coronal-dynamics, fast-wind) to cover the main intent of the observations, as loosely as possibly so that it does not exclude any of the intended sub-objectives. The name may not always be self-evident, but it is indicative of the observations. There are three SOOP names that do not follow the above rules (\emph{IS\_DEFAULT}, \emph{L\_IS\_STIX} and \emph{L\_IS\_SoloHI\_STIX}), these SOOPs describe coordinated synoptic observations with the instruments named in the SOOP names (\emph{IS} stands for in-situ). The different names of SOOPs in the current SAP release can be found in Table 1 and are detailed in the following sections (note that all SOOPs of this Table are clickable and point to the relevant sections). For details about the instruments and their operational modes,  see the individual instrument papers in this issue\footnote{\url{https://www.aanda.org/component/toc/?task=topic&id=1082}}. For details about each SOOP and how it is going to address the particular sub-objectives, see the SAP pages of the Solar Orbiter science website.

Lastly, these SOOP names play a key role in identifying the resulting cohesive scientific datasets to the user in the Solar Orbiter ARchive (SOAR)\footnote{\url{http://soar.esac.esa.int/soar/}}. Once they have already been downlinked, processed, and checked for calibration, all datasets from a given SOOP, from all participating instruments, will be linked and discoverable by searching for that SOOP name.

\begin{table}
\label{tab:soops}
\caption{List of Solar Orbiter Observing Plans (SOOPs). Every line is a clickable link to the respective SOOP section.}
\centering                          % used for centering table
\begin{tabular}{l}        % centered columns (4 columns)
\hline\hline                 % inserts double horizontal lines
\\
SOOP names \\    % table heading 
\\
\hline                        % inserts single horizontal line
 % \multirow{3}{*} 
\\ 
\hyperlink{isdefault}{IS\_DEFAULT}\\
\hyperlink{lisstix}{L\_IS\_STIX}\\
\hyperlink{lissolohistix}{L\_IS\_SoloHI\_STIX}\\
\hyperlink{coronalsynoptic}{L\_FULL\_LRES\_MCAD\_Coronal-Synoptic}\\
\hyperlink{probequadrature}{L\_FULL\_LRES\_MCAD\_Probe-Quadrature}\\
\hyperlink{cmeseps}{L\_FULL\_MRES\_MCAD\_CME-SEPs}\\
\hyperlink{magnfieldconfig}{L\_FULL\_HRES\_LCAD\_MagnFieldConfig}\\
\hyperlink{coronalheabundance}{L\_FULL\_HRES\_MCAD\_Coronal-He-Abundance}\\
\hyperlink{eruptionwatch}{L\_FULL\_HRES\_HCAD\_Eruption-Watch}\\
\hyperlink{coronaldynamics}{L\_FULL\_HRES\_HCAD\_Coronal-Dynamics}\\
\hyperlink{ballisticconnection}{L\_SMALL\_MRES\_MCAD\_Ballistic-Connection}\\
\hyperlink{connectionmosaic}{L\_SMALL\_MRES\_MCAD\_Connection-Mosaic}\\
\hyperlink{fastwind}{L\_SMALL\_HRES\_HCAD\_Fast-Wind}\\
\hyperlink{slowwindconnection}{L\_SMALL\_HRES\_HCAD\_Slow-Wind-Connection}\\
\hyperlink{farsideconnection}{L\_BOTH\_MRES\_MCAD\_Farside-Connection}\\
\hyperlink{poletopole}{L\_BOTH\_LRES\_MCAD\_Pole-to-Pole}\\
\hyperlink{flareseps}{L\_BOTH\_MRES\_MCAD\_Flare-SEPs}\\
\hyperlink{chboundaryexpansion}{L\_BOTH\_HRES\_LCAD\_CH-Boundary-Expansion}\\
\hyperlink{majorflare}{L\_BOTH\_HRES\_HCAD\_Major-Flare}\\
\hyperlink{transitioncorona}{R\_FULL\_LRES\_LCAD\_Transition-Corona}\\
\hyperlink{globalhelioseismology}{R\_FULL\_LRES\_HCAD\_Full-Disc-Helioseismology}\\
\hyperlink{localareahelioseismology}{R\_FULL\_LRES\_HCAD\_Local-Area-Helioseismology}\\
\hyperlink{densityfluctuations}{R\_FULL\_HRES\_HCAD\_Density-Fluctuations}\\
\hyperlink{arlongterm}{R\_SMALL\_MRES\_MCAD\_AR-Long-Term}\\
\hyperlink{compositionvsheight}{R\_SMALL\_HRES\_LCAD\_Composition-vs-Height}\\
\hyperlink{finescalestructure}{R\_SMALL\_HRES\_LCAD\_Fine-Scale-Structure}\\
\hyperlink{polarobservations}{R\_SMALL\_HRES\_MCAD\_Polar-Observations}\\
\hyperlink{atmosphericdynamicsstructure}{R\_SMALL\_HRES\_HCAD\_Atmospheric-Dynamics-Structure}\\
\hyperlink{ardynamics}{R\_SMALL\_HRES\_HCAD\_AR-Dynamics}\\
\hyperlink{pdfmosaic}{R\_SMALL\_HRES\_HCAD\_PDF-Mosaic}\\
\hyperlink{rsburst}{R\_SMALL\_HRES\_HCAD\_RS-Burst}\\
\hyperlink{wavestereoscopy}{R\_SMALL\_HRES\_HCAD\_Wave-Stereoscopy}\\
\hyperlink{ephemeral}{R\_SMALL\_HRES\_HCAD\_Ephemeral}\\
\hyperlink{granulationtracking}{R\_SMALL\_HRES\_HCAD\_Granulation-Tracking}\\
\hyperlink{nanoflares}{R\_BOTH\_HRES\_HCAD\_Nanoflares}\\
\hyperlink{filaments}{R\_BOTH\_HRES\_HCAD\_Filaments}\\
\\
\hline
    
\end{tabular}
\end{table}

\subsection{The in-situ baseline SOOP IS\_DEFAULT}
\mytarget{isdefault}{}

IS\_DEFAULT is the baseline SOOP involving all four in-situ instruments EPD (Energetic Particle Detector, see \citealt{Rodriguez2020a}), MAG (Magnetometer, see \citealt{Horbury2020}), RPW (Radio and Plasma Waves instrument, see \citealt{Maksimovic2020a}) and SWA (Solar Wind Analyser, see \citealt{Owen2020a}).
This will be the standard SOOP that will run during the whole cruise phase and during the nominal phase outside the RSWs. It will also run in combination with other SOOPs during the RSWs. It is designed to address all in-situ objectives for which no remote-sensing observations are needed.
Within the SOOP, the four instruments will be in normal or in burst mode, together or individually. For details regarding the coordinated campaigns of in-situ instruments, see \citet{Walsh2020}. 

This SOOP covers the operations needed to address all top-level goals regarding the solar wind, interplanetary magnetic fields, coronal mass ejections and energetic particles. For many of those, it is important to perform long-term observations with a good spatial coverage of the inner heliosphere during the different phases of the solar cycle. This is why the in-situ instruments will always be operating throughout the mission, within the operational constraints explained in \citet{Sanchez2020}. Special care has been taken for Electro-Magnetic Cleanliness (EMC) and it is guaranteed that the spacecraft will be EMC-quiet during at least $70\%$ of the time of the mission in order to acquire meaningful in-situ data (see \citealt{Sanchez2020} and \citealt{Garcia2020}).

Scheduling of different scientific objectives will have to take into account the best orbital opportunities. For instance, there are optimal geometrical configurations (such as radial alignments, alignments along the interplanetary magnetic field, and quadratures) between Solar Orbiter, Parker Solar Probe, BepiColombo, the Sun and Earth for which it will be particularly interesting to schedule burst modes (for the collaboration with other space and Earth assets, see \citealt{Velli2020a}). Whenever additional assets are used to improve our understanding of the various sub-objectives, these have been indicated in the SAP.

In all of the following SOOPs, the in-situ instruments are always on, at least in normal mode, and we will not explicitly refer to their roles if not specifically needed for the SOOP or if they are not different from this mode.

\subsection{L\_IS\_STIX: the in-situ default SOOP enhanced with X-ray observations}
\mytarget{lisstix}{}

For some objectives related to solar energetic particles (e.g. how they can be rapidly accelerated to high energies, seed particles provided by flares, accelerated electrons on short timescales to explain hard X-ray fluxes), it is important to have information about X-ray emission in addition to the EPD measurements. This will be achieved by coupling the previous \hyperlink{isdefault}{IS\_DEFAULT} SOOP to the normal mode of the STIX instrument (Spectrometer/Telescope for Imaging X-rays -- \citealt{Krucker2020a}). As per current plan, STIX is baselined to operate only during the RSWs, but executing this SOOP for longer periods, including outside the windows, would allow better statistics on observations of X-rays and energetic particles over the course of the solar cycle and across a range of heliocentric distances. 

\subsection{L\_IS\_SoloHI\_STIX: the L\_IS\_STIX SOOP enhanced with spatial context from SoloHI}
\mytarget{lissolohistix}{}

This SOOP aims to measure variability in gradual solar energetic particle events through the corona and the heliosphere. It is based on the previous \hyperlink{lisstix}{L\_IS\_STIX} SOOP, enhanced with spatial context from the heliospheric imager, SoloHI \citep{Howard2020a}. This is for instance important to determine the role of warped shock fronts, turbulence and inhomogeneities in general, solar wind turbulence. The SoloHI Shock, Turbulence or Synoptic modes will be used (see \citet{Howard2020a} for details) depending on the distance from the Sun (shock and turbulence modes are scheduled when the spacecraft is inside 0.4\,au). Some of the objectives require in-situ burst modes, whereas others would greatly benefit from coordination with Parker Solar Probe. Similarly, this SOOP will also be used to augment Parker Solar Probe observations.

\subsection{L\_FULL\_LRES\_MCAD\_Coronal-Synoptic}
\mytarget{coronalsynoptic}{}

The coronal synoptic SOOP L\_FULL\_LRES\_MCAD\_Coronal-Synoptic is designed to provide insight into the global coronal structure and facilitate the study of coronal mass ejections. When scheduled, it typically runs for an entire RSW (default duration of ten days). The imagers point to disc centre (no off-pointing to a different region of the Sun is needed), and the SOOP will make use of the in-situ and Metis instrument triggers \citep{Antonucci2020a}. It is important to note that many instruments have triggers in their onboard systems. The EUI has a flare trigger responding to intensity enhancements in EUI/FSI (Full Sun Imager); STIX has a flare trigger responding to flares in the X-ray; and Metis has a CME trigger that responds to enhancements in white light emission off-limb. The EPD instrument triggers to burst mode when the energetic particle fluxes in different energy ranges reach certain thresholds. RPW triggers to burst mode when it detects interplanetary shocks or type III bursts and also prompts enhanced data capture for SWA and MAG.

The whole payload participates in this SOOP. The Extreme Ultraviolet Imager (EUI, see \citealt{Rochus2020a}) makes use of the FSI at its synoptic mode with a cadence of around ten minutes (all SOOPs' cadences may be adjusted when scheduling them). The Metis coronagraph uses the GLOBAL mode and the CMEOBS special mode whenever triggered. The Photospheric and Helioseismic Imager (SO/PHI, see \citealt{Solanki2020a}) observes with the Full Disc Telescope (FDT) synoptic mode with a cadence of six hours. SoloHI performs normal observations (with Synoptic and Shock modes at perihelion). STIX and in-situ instruments are in normal or burst modes (scheduled or triggered).

Objectives that will be addressed with observations made during this SOOP include CME structure and evolution, how the Sun's magnetic fields link into space, heating in flaring loops vs heating in active regions, and properties of the magnetic field at high solar latitudes. This SOOP will also be used for tracking supergranules in the Doppler maps to obtain maps of the large scale flows, including rotation and meridional circulation at high latitudes. For this purpose, Dopplergrams will be used taken at a 6-hour cadence. The resolution needed is a few hundred pixels across the solar diameter. The same series of Dopplergrams will be used to study the evolution of convective modes and supergranulation waves.

This SOOP is specifically designed to be low-resource and can be run even under the most challenging telemetry constraints. In that sense, this SOOP describes the minimum of performance that is feasible using the remote-sensing payload during the RSWs.

\subsection{L\_FULL\_LRES\_MCAD\_Probe-Quadrature: specific SOOP for Parker Solar Probe quadratures}
\mytarget{probequadrature}{}

This SOOP is designed to study the corona while Parker Solar Probe is in quadrature with Solar Orbiter, that is the Probe-Sun-Orbiter angle is of the order of 90$^\circ$. The default duration of this SOOP is three days, the Solar Orbiter spacecraft points the remote-sensing instruments to solar disc centre and the in-situ triggers are enabled. SO/PHI and EUI provide context, while Metis and SoloHI are the main instruments of this SOOP as they will be providing imagery of the solar wind that is expected to be encountered by the Parker Solar Probe. Medium cadence is sufficient for this SOOP: SO/PHI will be using the FDT telescope in a 6-hours cadence, EUI will be using the synoptic mode of FSI, while Metis will be using its special COMET/PROBE mode with a cadence ranging between one and ten minutes depending on the heliocentric distance. SoloHI will also adapt its mode depending on the distance. 

\subsection{L\_FULL\_MRES\_MCAD\_CME-SEPs: Solar Energetic Particles accelerated by CMEs}
\mytarget{cmeseps}{}

In order to understand whether and how the in-situ properties of the solar energetic particles (SEPs) are linked to the coronal mass ejections (CMEs) and shocks, continuous observations with the in-situ payload are needed, together with SoloHI and Metis instruments observing the CMEs as they propagate out of the corona. The typical duration of this SOOP is ten days with a disc centre pointing and all in-situ and Metis triggers enabled. SoloHI will be observing in its Shock and Synoptic modes, Metis will be in the standard mode GLOBAL or in the specially designed mode CMEOBS. The FSI imager of EUI will be in synoptic mode with the default cadence of ten minutes or down to five minutes when possible for specific objectives. SO/PHI and SPICE (SPectral Imaging of the Coronal Environment -- \citealt{SpiceConsortium2020}) are not included in this SOOP. If possible, this SOOP will run when the Parker Solar Probe is within 0.25\,au and optimally to the east of Solar Orbiter, so that Probe is into the SoloHI field of view.

\subsection{L\_FULL\_HRES\_LCAD\_MagnFieldConfig: studying the large scale structure of the interplanetary magnetic field}
\mytarget{magnfieldconfig}{}

Solar Orbiter will acquire magnetograms of the Sun's polar regions with SO/PHI, while MAG measures the magnetic field in space at a range of locations, making more precise measurements of the reversal of the solar magnetic field and its effects on the heliosphere. This SOOP has been mainly designed to understand how the solar field reversal affects the coronal and heliospheric magnetic fields. For this purpose, the instruments will measure the polarity and the variation of large scale structures of the Sun's magnetic field in interplanetary space close to the Sun as it progresses from solar minimum towards maximum and the global field reverses. Even though part of this objective can be addressed with in-situ only measurements, full disc remote sensing observations are required to properly understand how the changes observed in situ are linked to the overall changes of the Sun's surface magnetic field. For this, long-term observations at synoptic modes are needed (for telemetry reasons), by targeting the full disc for both photospheric and coronal fields. Since such (low resolution) observations already exist from Earth, it is mostly of interest to observe when Solar Orbiter is at the far side of the Sun or during the extended periods when the spacecraft is moving towards/away from Earth, at either side of the Sun. In the latter case it would be useful to take observations that are regularly spaced in solar longitude rather than in time. This will also be repeated for different latitudes. Since in-situ MAG measurements are key for this objective, good statistics during EMC Quiet periods are required. SO/PHI's FDT will observe at its highest spatial resolution with a cadence of once or twice per day. EUI's FSI will also observe at its highest spatial resolution at the same cadence. Metis will observe with its synoptic programme for magnetic field structure (GLOBAL and/or LT-CONFIG modes). 

\subsection{L\_FULL\_HRES\_MCAD\_Coronal-He-Abundance: linking helium abundance from the corona to the solar wind}
\mytarget{coronalheabundance}{}

Compared to its value in the solar convective envelope, the helium abundance deduced from in-situ measurements of the fast and slow solar wind has long been known to be depleted relative to hydrogen, with occasional transient exceptions \citep{1998SSRv...85..291B}. In the slow solar wind, the degree of depletion has been shown to depend upon the wind speed and the level of solar activity \citep{2001GeoRL..28.2767A}. Measurements of the helium abundance in the corona, associated with measurements of the coronal outflow velocity, will provide evidence of the degree of correlation between wind speed and helium abundance and allow identification of the source regions of slow wind streams with different helium abundance. During
the mission phases when the spacecraft is at the closest perihelia of 0.28\,au, it will be able to continuously observe individual regions, free from projection complications, over longer periods than are possible from Earth orbit. During this reduced rotation of the Sun relative to the spacecraft, the intrinsic evolution of magnetic topology will be inferred from photospheric magnetograms and extrapolations and thus its influence on the wind parameters (such as wind outflow velocity and helium abundance) will be directly assessed. 

The abundance can be derived from simultaneous observations of the resonantly scattered component of singly ionised helium by EUI/FSI in its 30.4 nm channel and of that of neutral hydrogen by Metis in Ly$\alpha$ (121.6 nm). For EUI the FSI synoptic mode with 20 minutes cadence will be used and potentially the EUI occulter (the FSI aperture door has a position in which an occulting disc blocks the light coming from the solar disc, so that only the UV light coming from the low corona can reach the detector). For Metis, the modes MAGTOP and WIND will be used with a 20 minutes cadence for a duration of at least two hours in order to obtain global maps of neutral hydrogen Ly$\alpha$ intensity, electron density and the outflow velocity.
SPICE (Composition Mapping mode) will contribute a mapping of the near-surface elemental abundances, including that of helium, which constitutes a reference for establishing abundance variations in the wind. SO/PHI will contribute with a 6-hour cadence synoptic mode, providing data suitable for coronal magnetic field extrapolations. SoloHI will measure the solar wind speed above the potential source region in coordination with Metis. In those circumstances when connectivity could be established between the coronal regions observed by Metis and EUI/FSI and the in-situ plasma probed by Solar Orbiter, SWA will measure the $\alpha/p$ density ratio and composition. This SOOP will in principle be scheduled when the spacecraft is inside 0.45-0.5\,au, which is an optimal distance for the EUI/FSI occulter, or at perihelion for a reduced rotation of the Sun relative to the spacecraft. This SOOP would also benefit from coordination with DKIST (Cryo-NIRSP instrument) when Solar Orbiter is in conjunction or opposition with Earth. In particular, the Cryo-NIRSP spectrograph on DKIST in coronagraphic mode will be able to provide measurements of the He I 1083.0 nm line as well as context imaging and spectra in coronal lines in the infrared up to 1.5 solar radii.

\subsection{L\_FULL\_HRES\_HCAD\_Eruption-Watch}
\mytarget{eruptionwatch}

This SOOP is designed to observe eruptive events and contribute to the understanding of coronal mass ejection initiation. The full solar disc will be observed with highest spatial resolution available to the full disc telescopes and temporal cadence for a typical duration of one day. EUI will be in the Global Eruptive Event mode during the whole day, but it will prioritise the data of one or two events due to telemetry limitations (for example data corresponding to two hours of the day). Metis' standard mode GLOBAL will be used together with the CMEOBS mode for a duration of at least one hour, when triggered with a CME flag. SO/PHI's FDT will also be observing at the highest spatial resolution with a cadence of two to five minutes and a data selection will be needed as for EUI. SoloHI will perform a combination of the Shock and Synoptic modes, each for half of the total duration of the SOOP. SPICE will run a CME watch programme made of high (8\,min) and slow (22\,min) cadence sparce scans. These will have the largest possible field of view and will allow to sample CMEs initiation dynamics, dimming propagation and perform plasma diagnostics of density, velocity and temperature. STIX and the in-situ instruments will be in normal mode, with all triggers enabled. The EUI and SO/PHI internal memory capacities enable to run this SOOP during a maximum duration of one day. The preferred orientation for these observations is in quadrature with Earth (i.e. the angle between Earth-Sun-Orbiter being of the order of 90$^\circ$), so that L1 and Earth-based assets may measure the resulting ICME in situ in case it propagates towards the Earth. 

\subsection{L\_FULL\_HRES\_HCAD\_Coronal-Dynamics}
\mytarget{coronaldynamics}{}

This SOOP is aimed at observing structures in the outer corona and linking them to the heliosphere observed in-situ. The Metis and SoloHI instruments lead this SOOP, while the in-situ payload provides continuous observations. Synoptic support is provided from other full disc remote-sensing instruments (SO/PHI and EUI). The typical duration of this SOOP is one day. SoloHI operates in a combination of the high-cadence Turbulence mode and the Synoptic mode. Metis runs a generic mode such as WIND interleaved with FLUCTS for a typical duration of one hour per day. When this is run close to the perihelion, the disc centre pointing is preferred in order for Metis to be able to operate. When off-pointing to a limb active region is possible, SPICE can optionally contribute with the `CME Watch limb' mode. Objectives that will be covered with this SOOP are tracing of streamer blobs and other structures through the outer corona and the heliosphere, study of the structure and evolution of streamers, coronal shocks and associated heating, dissipation and acceleration mechanisms. When possible, radial alignments or quadratures with Parker Solar Probe and/or Earth assets would be very beneficial. 

\subsection{L\_SMALL\_MRES\_MCAD\_Ballistic-Connection}
\mytarget{ballisticconnection}{}

This SOOP also addresses the objectives of tracing streamer blobs or other structures through the outer corona. The difference to the previous SOOP is mainly that off-pointing is needed to the region on the Sun that is most likely connected ballistically to the spacecraft, according to the solar wind models. In this case, Metis cannot always be used. This SOOP will run in medium resolution and cadence for a typical duration of three days. EUI/FSI will be in Synoptic mode, while EUI/HRI (High Resolution Imagers) will observe in Coronal Hole mode at a cadence of roughly 15 minutes, adjusted to be equal to the cadence of SO/PHI's High Resolution Telescope (HRT). SoloHI will use its nominal synoptic perihelion programme. SPICE composition and dynamics modes will be interleaved. The in-situ instruments will be in normal mode with the regular scheduled bursts and the possibility of triggered burst. 

\subsection{L\_SMALL\_MRES\_MCAD\_Connection-Mosaic}
\mytarget{connectionmosaic}{}

This SOOP is designed to allow the high-resolution remote-sensing observations to cover a wider area than normal, in particular SPICE. This SOOP is to be used in particular with a mosaic of spacecraft pointings (see \citealt{SpiceConsortium2020} in this issue for details), though it does not necessarily need to be. Alternatively it could be used when the spacecraft points, for a limited amount of time (few hours typically), to the most likely connectivity point, within a mainly solar disc-centred observation period. The SPICE instrument, which needs 30 minutes to produce a composition map, is the priority instrument driving this SOOP with a typical duration of three hours. The details on the slit size, exposure time, number of positions, field of view, number of repetitions and the lines used can be found on the SAP pages of the Solar Orbiter science website for each objective. SO/PHI/HRT and EUI/HRI will be producing two to three images for each SPICE map, which corresponds to a typical cadence of 10-15 minutes. Synoptic observations will also be provided by EUI/FSI. The Metis instrument will in general be in safe mode with its door closed, unless the spacecraft is far enough from the Sun (see \citet{Antonucci2020a} for details about the operational constraints of Metis). In particular, the door of the Metis instrument must be closed when the spacecraft is pointing off Sun centre, called `off-pointing', since the coronagraph is designed specifically to operate in Sun-centred position. However the door can be open during off-pointing if at the same time the spacecraft is far enough from the Sun, approximately farther than 0.55\,au. This SOOP covers objectives related to the connection of the plasma observed in situ to the features observed on the solar disc. For example, to trace streamer blobs (as already described in SOOPs \hyperlink{ballisticconnection}{L\_SMALL\_MRES\_MCAD\_Ballistic-Connection} and \hyperlink{coronaldynamics}{L\_FULL\_HRES\_HCAD\_Coronal-Dynamics}), a north-south mosaic will be performed (thus eliminating latitudinal uncertainty), centred on the most likely connection point as defined by state-of-the-art models (see \citealt{Rouillard2020a}). This objective can equally benefit from Parker Solar Probe observations as well as observations from Earth (e.g. when in quadrature with Earth we can have SoloHI images of blobs directed to Earth or images taken from Earth of blobs that are expected to reach Solar Orbiter). Another objective that will be addressed with this SOOP is the identification of the possible sources of the slow solar wind, by doing, for example, mosaics of a larger area around an active region, again using modelling to identify the best possible candidate. This SOOP is expected to run many times in different phases of the solar cycle and near perihelion for a maximum chance of linking the plasma to its source regions.

\subsection{L\_SMALL\_HRES\_HCAD\_Fast-Wind}
\mytarget{fastwind}{}

To identify the sources of the fast solar wind, it is not sufficient to perform a remote-sensing characterisation of the corona, for example, by pointing to the centre of a coronal hole and its boundaries. It is also important to observe the fast wind in situ and link it back to its origin. In order for the in-situ plasma to be likely connected to the remotely-observed source regions, a coronal hole needs to be chosen, based on earlier observational evidence and modelling. Such a connected coronal hole is expected to be on average near the west limb and it needs to be observed from a location close to the Sun and preferably during high-latitude windows. This will be done during different parts of the orbit. An example could be to observe during a window that precedes perihelion to obtain the overall context of the solar disc and then observe during the perihelion for connectivity. Polar coronal holes will also be observed with Metis from the equatorial plane (no need for high-latitude observing windows, since it can be observed in the plane-of-sky above the hole). Low-latitude coronal holes will also be observed for intermediate speed outflow. Even if this has a lower priority (in terms of this specific science goal), it will not be neglected since this is the typical solar wind observed at Earth. The minimum and declining phase of the solar activity cycle is preferable for observing polar coronal holes. Low-latitude coronal holes can be observed at any phase of the cycle, but the probability to be at the right longitude at perihelion is low.

The required observations include EUI/HRI in Coronal Hole mode with one minute cadence, during one to two hours and 12 hours at lower cadence, SO/PHI observes at high resolution at one min cadence, while SPICE provides FIP (First Ionisation Potential) bias and velocity maps running for several hours three times per day. The in-situ instruments will be in normal mode for the connection and burst modes for more details. MAG Burst mode is required for ion cyclotron wave identification to distinguish from different acceleration and heating mechanisms as well as small scale changes, for example, jet-like features. 3-D distributions are needed at 1\,s scales in order to determine the properties inside and outside these features and relative heating efficiency for protons and alphas.

The possible remote-sensing targets will include wide regions of well extended coronal holes, as well as smaller regions for focusing on the different candidate sources of the fast wind, such as small, cool coronal loops, open magnetic funnels at the base of coronal holes, spicules with short lifetimes and macro-spicules within coronal holes, polar plumes, interplume regions, coronal hole boundaries etc. 

\subsection{L\_SMALL\_HRES\_HCAD\_Slow-Wind-Connection}
\mytarget{slowwindconnection}{}

Identifying the sources of the slow solar wind is expected to be one of the most challenging scientific goals of Solar Orbiter. As for all wind origin objectives, it is required to compare the elemental composition, temperatures and charge states of coronal features to the in-situ ones, adopting different strategies for different kinds of structures (such as helmet streamers, loops near active regions, streamer cores, edges of active regions etc.). This SOOP aims at catching with the remote-sensing instruments the dynamics of an open-closed field boundary, while the wind associated with such a boundary would then be measured in situ. High-resolution remote-sensing observations are required to catch the dynamics and since specific target pointing is needed, the best available models will be used for predicting the best candidate region to which the spacecraft is likely connected. A typical duration for this SOOP is three days. EUI Coronal Hole mode will be used at a cadence of one minute together with the synoptic mode of FSI. SO/PHI will acquire regularly spaced HRT data at medium to high resolution (600\,s default cadence, but one hour cadence will also be marginally sufficient for studying interchange reconnection at high resolution). SO/PHI will also make use of the Low-Latency magnetograms \citep{Solanki2020a} in order to better identify the most interesting periods that would be selected for downlink. SPICE is central to this objective using a combination of the Dynamics mode and of Composition Mapping rasters. The raster area will be optimised to make sure that the open-closed field boundaries are captured and the lines will be chosen according to the type of target. 

\subsection{L\_BOTH\_MRES\_MCAD\_Farside-Connection}
\mytarget{farsideconnection}{}

Just as for the SOOP \hyperlink{slowwindconnection}{L\_SMALL\_HRES\_HCAD\_Slow-Wind-Connection}, the aim is to observe the region to which the spacecraft is likely to be connected and address all connectivity science goals, both for fast and slow solar wind sources. When the spacecraft is at the far side of the Sun as seen from Earth, additional imagery of the full solar disc is needed with the SO/PHI/FDT synoptic programme at low cadence. When possible, a higher spatial resolution is preferred. All other instruments will follow the same strategy as in \hyperlink{slowwindconnection}{L\_SMALL\_HRES\_HCAD\_Slow-Wind-Connection}.

\subsection{L\_BOTH\_LRES\_MCAD\_Pole-to-Pole}
\mytarget{poletopole}{}

This SOOP is designed to be used as a whole or half-orbit synoptic campaign that scans the Sun from high latitudes in one hemisphere to the other. For that reason, it will mainly be used later in the nominal mission phase when the spacecraft reaches inclinations of at least 15 degrees. This SOOP resembles very much the coronal synoptic \hyperlink{coronalsynoptic}{L\_FULL\_LRES\_MCAD\_Coronal-Synoptic}, but this time SPICE is necessary as well.
The default duration of this SOOP is ten days with the imagers pointing mainly to the disc centre. EUI makes use of the FSI telescope in its synoptic mode with a cadence around ten minutes. Metis uses the GLOBAL or LT-CONFIG modes to observe large scale coronal structures and the CMEOBS special mode whenever triggered by a CME detection. SO/PHI observes with the FDT telescope synoptic mode, but HRT will also be used at a medium cadence especially at higher latitudes for polar magnetic field observations. SoloHI performs synoptic observations. STIX and in-situ instruments are in normal or burst modes (scheduled or triggered). SPICE scans many latitudes performing a Composition Mapping raster for mapping the whole area, followed by multiple instances of the CME Watch mode. 

\subsection{L\_BOTH\_MRES\_MCAD\_Flare-SEPs}
\mytarget{flareseps}{}

This SOOP is aimed at understanding the properties and dynamics of solar energetic particles (SEPs) in relation to flare events. EUI and STIX are leading this SOOP, while the in-situ payload provides continuous observations. Synoptic support from other full disc remote-sensing instruments is provided. Disc centre pointing is preferred especially since the SEP events are rather rare and it is not feasible to know which particular region should be observed. Because of the difficulty predicting the events, it is preferable not to observe with high-resolution or high-cadence with EUI and SO/PHI (five to ten minutes cadence is considered to be sufficient), except when such observations (at one minute cadence) are triggered by a STIX flag whenever the flare happens to be inside the EUI/HRI field of view. Metis special mode CMEOBS can also be triggered by a CME flag and observe for a minimum time of one hour with a cadence of one minute.

\subsection{L\_BOTH\_HRES\_LCAD\_CH-Boundary-Expansion}
\mytarget{chboundaryexpansion}{}

This SOOP is similar to \hyperlink{slowwindconnection}{L\_SMALL\_HRES\_HCAD\_Slow-Wind-Connection}, but it specifically aims to study over-expanded coronal holes boundaries as possible sources of the slow solar wind. This requires a different SO/PHI mode and different SPICE observations. Regularly spaced SO/PHI/FDT images are required with a cadence of six hours throughout a remote-sensing window. For SPICE, the raster area will be optimised to make sure the open-closed field boundary is captured. For this reason, six rasters are planned at perihelion at highest possible spatial resolution. A full north-south raster will only be done for extended holes, otherwise it is enough to point at the boundaries. For stable structures, three to four days of standard observations are planned together with one day of mosaic observations.

\subsection{L\_BOTH\_HRES\_HCAD\_Major-Flare}
\mytarget{majorflare}{}

This SOOP aims to perform high-resolution and high-cadence observations of a major flare to study the event in unprecedented detail, ideally at perihelion and pointing to the most promising region for a typical duration of four days. The first instance of this SOOP is planned in the ascending phase of the solar cycle. It is proposed to repeat this SOOP until a major flare is successfully captured at high cadence and high spatial resolution. A special opportunity could arise when Solar Orbiter and Earth are both connected to the same, promising Active Region (either magnetically or ballistically). It will potentially be combined with observations from DKIST and other Earth-based (GST, SST, Gregor etc.) or near-Earth observatories and instruments. EUI/HRI observes in highest spatial resolution with 1\,s cadence and EUI/FSI at 1\,min cadence. To avoid limiting the telemetry for the entire orbit, the triggers allow to only select the most interesting hour for downlink. The FSI images allow to study potential EUV waves associated with the flare. STIX observes in its normal mode. SPICE is in Dynamics mode with the raster centred at the centre of EUI. A SPICE CME watch mode could be added after a number of repeats of this SOOP. SO/PHI/FDT observations are used for context and extrapolations. SoloHI in normal mode provides heliospheric imaging of a potential CME associated with the flare.

\subsection{R\_FULL\_LRES\_LCAD\_Transition-Corona}
\mytarget{transitioncorona}{}

The large scale structure of the low corona is determined by active regions, filament channels and coronal holes whose presence and location have a clear evolution over the solar cycle. Higher up in the corona, active regions and filaments are over-arched by pseudo-streamers and streamers that fade into the heliospheric plasma sheet. This intermediate region or transition corona between about one and three solar radii above the limb has not to date been well studied as it corresponds to the field-of-view gap between most EUV imagers and coronagraphs. Coronal/heliospheric simulation codes such as ENLIL \citep{1999JGR...104..483O} or EUHFORIA \citep{2018JSWSC...8A..35P} typically bypass the complexity in this region completely by employing ad hoc empirical laws to obtain `coronal' boundary conditions at 0.1\,au. When trying to determine the coronal footpoint of features observed by the in-situ instruments, the transition corona introduces significant uncertainty as the Parker-spiral type of mapping has to be connected to magnetic field extrapolations from the photosphere. Deciphering this connection is thus important for Solar Orbiter connection science, but the transition corona itself might also harbour interesting features. The purpose of this SOOP is to obtain a full-Sun structure of the transition corona involving in first instance SO/PHI/FDT (for magnetic extrapolations), EUI/FSI (optionally with occulter) and Metis. The basic observation unit is one full-disc image set per day from all three telescopes or ideally in fixed longitude intervals. This basic observation unit has to be repeated as many subsequent days as possible as to obtain a full-Sun 360 degree coverage. This will be achieved by combining three RSWs, by joint observations from Earth-based instruments and — if needed — observations outside of the remote sensing windows will be considered too (see Section 6). The 360 degree total observation will be done at least once near solar minimum and once near solar maximum. This SOOP does not require a perihelion observation point for the highest resolution nor any off-pointings. 

\subsection{R\_FULL\_LRES\_HCAD\_Full-Disc-Helioseismology}
\mytarget{globalhelioseismology}{}

This SOOP is designed for all helioseismology techniques using full-disc data from the SO/PHI FDT. Solar Orbiter will provide the first opportunity to implement helioseismology to probe flows and structural heterogeneities deep in the Sun as well as active regions on the far side. Combining Solar Orbiter observations with ground- or space-based helioseismic observations from 1\,AU (e.g. GONG, SDO, Lagrange) will open new windows into the Sun \citep{Loeptien2015}. Looking at the Sun from two distinct viewing angles will increase the observed fraction of the Sun's surface and will benefit global and local helioseismology techniques (e.g. \citealt{2016A&A...592A.106R}), because the modes of oscillation will be easier to disentangle due to the reduction of spatial leaks. The signal-to-noise will increase and the reliability of the measured mode frequencies will improve due to the reduced reliance on details of the modelling.

To fulfil the full-disc helioseismology objectives, the SO/PHI FDT line-of-sight velocity images are required at one minute cadence, together with some intensity and magnetic field images at much lower cadence for context information. In order to save telemetry, the spatial resolution may be reduced by a variable amount (depending on distance to the Sun and centre-to-limb position) and images could be compressed by binning to 2$\times$2 (corresponding to a resolution of 3\,Mm when the spacecraft is at perihelion) or cropping when observing far away from the Sun. Higher compression may possibly be acceptable for far side imaging. For deep focusing, this SOOP will run for several days (e.g. three days). For far side imaging to detect modes passing through the solar core, observations are needed to be as long as possible, of the order of 60 days. Such long observations are not yet part of the science plan and their feasibility needs to be assessed taking into account the operational limitations.

Two spacecraft (or the Earth plus a spacecraft) not only allows for increased spatial coverage, it also allows one to study mode physics by measuring two of the three components of the velocity, instead of only one. This allows one to investigate the phase difference between and the relative amplitude of the components as a function of position on the solar disc, which contributes to centre-to-limb systematics. These measurements can be compared to numerical simulations to learn about the physics of acoustic modes near the surface, provide a unique probe of the near-surface convection. By combining three vantage points (e.g. Solar Orbiter, GONG or SDO, and Lagrange at L5), it may even be possible to further constrain the three velocity components of the oscillations and the convection.

The differential rotation and the meridional circulation can also be measured using global-mode techniques by studying how the mode eigenfunctions are distorted at the surface \citep{2011ApJ...734...97S, 2013ApJ...778L..38S, Woodard2013}. These mode-coupling techniques will greatly benefit from an improved understanding of mode physics.

Techniques of helioseismic holography will also be applied to image the Sun's three dimensional structure by focusing acoustic waves at target locations in the solar interior. One of the most impressive applications of helioseismic holography is the imaging of active regions on the far side of the Sun, that is on the hemisphere facing away from Earth at a given time. Recent theoretical efforts have led to improvements of the method \citep{Gizon2018} and active regions of average sizes can now be imaged with only a few days of observations. Observations from Solar Orbiter will allow helioseismic holography to provide improved measurements of the remaining far side. In addition, SO/PHI magnetograms will help calibrate and validate the measurements of the far side made using holography based on data taken on the front side (from L1, the Earth, or L5), for improved space weather predictions.

Large-scale flows, including the meridional flow, can be constrained using time-distance helioseismology, which consists of measuring the time of acoustic waves travelling between pairs of points on the surface through the solar interior. Inversions of travel times have led to different answers for the deep meridional flow. These differences may result from different instrument systematic errors, the calibration of the observations, and/or different assumptions in the data analysis. An additional data source (SO/PHI FDT) will be important to sort out the current disagreement between MDI, GONG and HMI, to understand the systematics, and to improve the calibration of the observations. Furthermore, stereoscopic helioseismology will allow us to consider acoustic ray paths which reach deep in the convection zone or propagate through the polar regions (multiple skips), see \citet{Loeptien2015}. Another application of time-distance helioseismology is the study of solar Rossby waves, which play an important role in the dynamics of the convection zone. The Rossby waves with the longest wavelengths are particularly interesting as they have maximum sensitivity near the base of the convection zone; they will be much easier to detect with extended spatial coverage from L5.

\subsection{R\_FULL\_LRES\_HCAD\_Local-Area-Helioseismology}
\mytarget{localareahelioseismology}{}

This SOOP has several scientific objectives, all revolving around the use of SO/PHI/HRT data for helioseismology \citep{Loeptien2015}. These include studying large scale flows and rotating convection in the high latitude regions, as well as the structure and dynamics of active regions. Mode physics and the study of systematics in local helioseismology using multiple views are additional objectives.

The observations needed are Dopplergrams from the HRT at full cadence (1\,min) and at a high spatial resolution (at most 1\,Mm on the Sun). The time series should be continuous and should be as long as possible (from days to months). High spatial resolution is required to observe the seismic waves with the shortest wavelengths; for example, the quarter wavelength of the f mode is 1.2\,Mm at 3\,mHz. Seismic modes f and p$_1$ through p$_4$ are used to study the dynamics of the surface layers at spatial scales down to supergranulation \citep{Gizon2010}. The Doppler data will be supplemented by an intensity image and a magnetogram every hour to provide context information.

The full resolution of the CCD will be needed near the limb in the high-latitude regions to take account of foreshortening, while it may be degraded towards disc centre. In addition, lossy compression techniques can be applied to further reduce the telemetry \citep{Loeptien2014}.

\subsection{R\_FULL\_HRES\_HCAD\_Density-Fluctuations}
\mytarget{densityfluctuations}{}

This SOOP involves only remote-sensing instruments and is designed to study the density fluctuations in the extended corona as a function of the outflow velocity of the solar wind. Metis and SoloHI are the primary instruments in this SOOP with the Metis FLUCTS mode for one hour (with cadence ranging from 1-20s), then MAGTOP (5-20 minutes cadence) for several hours to several days. Preferably the SOOP would be repeated for eight hours during the two following RSWs. SO/PHI/FDT will observe at 6-hour cadence to provide context. Long exposures for EUI/FSI synoptic mode are needed to get good signal-to-noise ratio where it overlaps with Metis. SPICE will optionally participate before and after the main observations, while off-pointing to an active region at the limb. 

\subsection{R\_SMALL\_MRES\_MCAD\_AR-Long-Term}
\mytarget{arlongterm}{}

This remote-sensing SOOP will study the decay process of active regions. It needs to run for a typical duration of 15 days, centred on perihelion, continuously pointing to an active region, to fully catch the evolution and dispersion of that region. This SOOP makes the best use of reduced rotation of the Sun relative to spacecraft, following the active region for as long as possible without the disadvantages of too much foreshortening and changes in viewing angle. SO/PHI is the primary instrument in this SOOP with full field-of-view observations (FDT) at a cadence of ten minutes. SPICE Dynamics mode will be used. EUI will match the field of view, resolution and cadence of SO/PHI. Metis will potentially provide context data before and after this SOOP. 

\subsection{R\_SMALL\_HRES\_LCAD\_Composition-vs-Height}
\mytarget{compositionvsheight}{}

This SOOP will map the abundance of minor ions as a function of height in the corona to distinguish between slow and fast solar wind. The observation target can be a boundary of a streamer or an active region at the limb. SPICE is the primary instrument in this SOOP with the Composition Mapping mode. EUI will provide context at higher cadence than SPICE in order to interpret the SPICE composition map. SO/PHI is needed for context magnetic field, but mainly before or after the SPICE observations since the target will be at the solar limb. Since Metis cannot participate at limb pointing (because it cannot off-point when Solar Orbiter is inside 0.55\,au), it will provide context observations before and after the main observations. A typical duration of this SOOP is of the order of several hours, run twice. This can be done at any remote-sensing window, but perihelion is preferred for active region observations or a distance greater than 0.55\,au for streamer observations (since this is the distance beyond which Metis will observe even when the spacecraft is off-pointing). 

\subsection{R\_SMALL\_HRES\_LCAD\_Fine-Scale-Structure}
\mytarget{finescalestructure}{}

This SOOP studies the finest scales of active regions and other solar features. This SOOP needs to be at the highest spatial resolution, but low cadence is sufficient since, in this case, the dynamics are not important (this is covered by other SOOPs). A perihelion window is preferred for these observations. EUI/HRI will use the Quiet Sun and Active Region modes at a cadence of ten minutes and at the highest resolution. SO/PHI/HRT would match the cadence and resolution of EUI. Depending on the science goal (e.g. for waves and/or temperature structure discrimination), SPICE will be in one of its high resolution modes, e.g. Dynamics. A typical duration is 12 hours. Metis will take part to provide off-limb observations when close to the Sun, e.g. for studying plumes. 

\subsection{R\_SMALL\_HRES\_MCAD\_Polar-Observations}
\mytarget{polarobservations}{}

This SOOP is designed to address the objectives relevant to the polar magnetic field, which do not necessarily rely on the highest resolution and cadence of SO/PHI/HRT nor on all five physical parameters that SO/PHI can return. The typical cadence for SO/PHI and EUI is two to five minutes. SoloHI and STIX observe using their nominal modes, while SPICE uses a combination of composition and dynamics modes. Metis will have its door closed for safety reasons due to off-pointing. 

\subsection{R\_SMALL\_HRES\_HCAD\_Atmospheric-Dynamics-Structure}
\mytarget{atmosphericdynamicsstructure}{}

This SOOP is designed to study the fine structure of the atmosphere. It is a burst mode of one hour for EUI (HRI at 1-30 s cadence), SO/PHI (HRT at one minute cadence) and SPICE. The instruments Metis, SoloHI and STIX do not participate in this SOOP. Many objectives will need this SOOP such as the full characterisation of photospheric magnetic fields and flux removal, high-latitude magnetic field structures, flux appearance modes and interaction in Quiet Sun, granulation and oscillations, limb stereoscopy of magnetic fields etc. This SOOP also serves to probe the coupling of the solar atmosphere by the magnetic field from the photosphere to the corona and how the photospheric field drives chromospheric and coronal dynamics and heating. 
Due to the very high spatial resolution and high time cadence, this SOOP is very demanding from the telemetry point of view. For this reason, it will only be run for short times and only when the onboard memory has enough storage capability available, preferentially in a perihelion window.

\subsection{R\_SMALL\_HRES\_HCAD\_AR-Dynamics}
\mytarget{ardynamics}{}

This SOOP will study the dynamics of a complex active region and its link to the ${}^3$He-rich SEPs production. It can also track a region for exploring the formation of an active region. It will make use of the highest resolution capabilities without binning. EUI will run its Active Region mode in coordination with SO/PHI/HRT. Due to telemetry limitations, only one hour of these data around the event will be downlinked.
STIX will have its triggers active. EPD is also needed for detecting the  ${}^3$He-rich SEP events (but this is not part of the SOOP, since all in-situ instruments will always be on in parallel with any other SOOPs). 

\subsection{R\_SMALL\_HRES\_HCAD\_PDF-Mosaic}
\mytarget{pdfmosaic}{}

This SOOP observes the probability density function of magnetic elements on the solar surface. It scans the solar radius with a mosaic made up of three to four different positions from the equator to a pole and from the disc centre to the east and west limb for calibration purposes. SPICE would require 30 minutes at each dwell position with at least ten images per dwell taken by SO/PHI/HRT and EUI/HRI giving information of how the corona changes during a SPICE scan.

\subsection{R\_SMALL\_HRES\_HCAD\_RS-Burst}
\mytarget{rsburst}{}

This SOOP describes a coordinated observation of high-resolution remote-sensing instruments (EUI, SO/PHI and SPICE), running at highest spatial resolution and variable, but high cadence, for a short period of time (of the order of ten minutes). As a planning scenario, it is proposed to run this SOOP at every perihelion window where extra telemetry is available, or where the campaign would fit without sacrificing too much of the rest of the orbit. This SOOP will be run for different targets, also at plain disc centre, as it is aimed to discover new physical phenomena and compare high cadence dynamics in all kinds of solar regions. In this case, running of Metis FLUCTS/TBF observing modes is very useful, to investigate coronal density fluctuations at high cadence (1\,s - 20\,s). EUI will use various HRI modes, including its Discovery mode designed to discover solar variations with periods less than ten seconds. 

\subsection{R\_SMALL\_HRES\_HCAD\_Wave-Stereoscopy}
\mytarget{wavestereoscopy}{}

The scientific aim of this SOOP is to characterise the properties of waves in the photosphere and their coupling with the atmosphere. Waves are one mechanism for transferring energy from the photosphere to the chromosphere and corona. The line-of-sight component of the wave velocity can be determined at different heights in the atmosphere by observing Doppler shifts. Determining the horizontal velocity has previously relied on using correlation tracking of intensity variations, based on the questionable assumption that the changes in location of the brightness fluctuations reflect the actual velocity. Solar Orbiter's orbit and capability to measure Doppler velocities, in conjunction with existing and upcoming ground-based or near-Earth observatories, offers the unique chance to directly measure two components of the velocity field using the Doppler effect.

High-resolution co-temporal measurements including Doppler velocity maps from SO/PHI as well as ground and near-Earth observatories are required. In particular the ground-based and near-Earth observations should include high resolution Doppler images in the same line (with a higher cadence than that of Solar Orbiter), as well as lines sampling different heights of the atmosphere. Co-observation with IRIS would be desirable. During the observing period, the Earth-Sun-Solar Orbiter angle will be between 30 and 60 degrees‚ a range which represents a compromise between determining the two components of the velocity field and allowing magnetic features, which can act as wave guides, to be partially resolved.

To understand the connection between the different heights, the observations are best performed at the centre of the disc as observed from Earth (where observations over different wavelengths are possible). Because the achievable cadence will be higher on ground than with SO/PHI, it is preferable to select targets which are closer to disc centre as seen from Earth and at larger heliographic angles as seen from Solar Orbiter. The highest possible cadence is desirable, and a shorter time series (down to 30 minutes of Solar Orbiter observations) would still allow the scientific objectives to be met. The ground-based and near-Earth observations should be made for a period of 90 minutes centred on the 30 minute Solar Orbiter observations. However, in order to guarantee good seeing at the coordinating ground-based observing facility (e.g. DKIST, GST, SST, NVST, Gregor) a continuous high-cadence observation period of several hours is required. High resolution context magnetic maps from Solar Orbiter immediately before and after the 30 minute observing window are required to provide context and aid co-alignment. A second observational campaign of an area 45$^\circ$ from disc centre, with an Earth-Sun-SO angle of 90$^\circ$, would be desirable. 

This SOOP will also be used to study the physics of helioseismic waves, in particular how the waves interact with the granulation and how they behave with height and viewing angle. Observations from various combinations of viewing angles is particularly important to determine the relationship between different components of the velocity. This will in turn allow us to validate the results of theoretical calculations and simulations and thereby increase our confidence in our measurements of large scale flows. Another objective is to study the magneto-acoustic waves in strongly magnetised regions.

\subsection{R\_SMALL\_HRES\_HCAD\_Ephemeral}
\mytarget{ephemeral}{}

This SOOP is designed to study the emergence, diffusion and decay of ephemeral regions near the poles and below high-latitude coronal holes. SO/PHI/HRT will observe half of the field of view with a cadence of one to two minutes at the highest resolution. EUI will match the field of view and cadence of SO/PHI. SPICE will produce composition maps. This SOOP will be carried out during high-latitude RSWs.

\subsection{R\_SMALL\_HRES\_HCAD\_Granulation-Tracking}
\mytarget{granulationtracking}{}

The objective of this SOOP is to determine horizontal flows at the surface at intermediate and large spatial scales (down to supergranulation). This is done by cross-correlating pairs of intensity images (SO/PHI HRT) taken one minute apart to determine the advection of granules. Since granules have a lifetime of $\sim$\,min, the pairs of images may be transmitted only once every around ten minutes without leading to a large increase in the noise. Additional data compression techniques have been proposed to reduce telemetry \citep{Loeptien2016}. The correlation tracking technique will also be applied to magnetic features using line-of-sight magnetograms, possibly at a reduced spatial resolution.

Surface flows deduced from feature tracking techniques are particularly interesting to study the dynamics at high latitudes, where the Coriolis force plays an important role. They are also key to calibrate and validate helioseismic methods.

\subsection{R\_BOTH\_HRES\_HCAD\_Nanoflares}
\mytarget{nanoflares}{}

This SOOP aims to detect and determine systematic differences between nanoflares (e.g. cadence, strength, location, magnetic field configuration) in different regions, such as the quiet Sun, active regions and coronal holes, but also any other target region in-between. Determining differences in nanoflare characteristics is crucial for a full understanding of the role of nanoflares in coronal heating. In order to study the latitude distribution of nanoflares, this SOOP will be run at different times. The default SOOP duration is one hour for telemetry reasons since EUI/HRI is at high-cadence (one second) and EUI/FSI at medium cadence (one minute). If longer observation time is needed, this SOOP will be run with up to 10\,s cadence. SO/PHI/HRT observations are used for context and information of possible changes of the magnetic field orientation and amount of magnetic flux in high resolution. Metis FLUCTS mode is equally needed with 1\,s cadence and can be followed by the mode observing total brightness fluctuations with a typical cadence of 20\,s. SPICE observes with an alternation of its Dynamics and Waves modes at 5\,s cadence. STIX is in normal mode.

\subsection{R\_BOTH\_HRES\_HCAD\_Filaments}
\mytarget{filaments}{}

This SOOP is intended to study the structure and dynamics of filaments at high spatial resolution. It supports both high and low cadence, depending on structural or dynamic aims. For the highest spatial sampling it is recommended for use during perihelion and will potentially be used in quadrature with Earth for coordinated stereoscopic observations with DKIST and other Earth-based (e.g. SST, GST, NVST, Gregor, ALMA) or near-Earth observatories and instruments. The default SOOP duration is one hour for telemetry reasons since EUI/HRI will operate at medium cadence (10-60\,s), but other instruments will observe for a longer period, such as Metis when off-pointing is not needed. SO/PHI/FDT observations will be taken for context and to facilitate extrapolations and SO/PHI/HRT at low cadence for magnetic stereoscopy. SPICE observes in the Dynamics mode with the raster centred at the filament.

It is important to remember at this point that the above list contains the SOOPs that were defined before launch. However, this SOOP library is expected to be a developing, living entity and members of the community are welcome to suggest new SOOPs or variants as new knowledge becomes available.
%__________________________________________________________________

\section{Observing strategy}

One of the main reasons for mission level planning is to address the science goals of the mission in an optimal way. More importantly, we need to make sure that if a unique opportunity exists for a specific science goal at a given date or configuration, this goal will be given priority. From this point of view, priorities have to be given, but they should not be based on the importance of the science objective (which is sometimes difficult to judge depending on the interests of the science community that cannot be anticipated many years in advance), but on many other circumstances that are detailed below. We built the plan without prioritising the objectives -- all four goals and their sub-objectives are equally important --, but we defined criteria for distributing the different SOOPs throughout the nominal and extended mission timelines. The overarching goal of the mission is to connect remote-sensing observations of the solar disc and corona with in-situ measurements (e.g. for CMEs, solar wind). This takes priority over everything else, whenever possible. These goals are usually called `connection-', `connectivity-' or `linkage-' science and are thoroughly detailed in \cite{Mueller2020a}. Except for this, the criteria that we use are detailed below in order of implementation. The implementation itself (i.e. the details of which SOOP is scheduled when) is out of the scope of the present paper. However we describe an implementation example in the next section for illustrative purposes only.

\subsection{Criterion 1: Best resolution remote-sensing data at different perihelia through the mission}

We schedule a remote-sensing burst SOOP (\hyperlink{rsburst}{R\_SMALL\_HRES\_HCAD\_RS-burst}) whenever we have a perihelion at a time of large telemetry downlink availability, that is close to Earth or when Solar Orbiter is moving towards Earth. We will aim at different types of targets or even plain disc centre to discover new physical phenomena in Solar Orbiter's highest cadence data, also possibly in unexpected locations. This campaign would serve several science goals that need very high cadence, need perihelion observations, and are aiming at different types of regions. Even if off-pointing is not possible, this campaign will still be useful to be run on the Sun-disc centre region. The SOOP under consideration is telemetry demanding, so we need large telemetry downlink availability at the time of the perihelion or right afterwards. Alternatively, for some science objectives, such as the study of the effects of energetic particles propagating downward in the chromosphere, it will be beneficial to schedule a few short sequences of the above SOOP in a RSW (or a series of RSWs), to enhance the chances of catching energetic particles instead of dedicating all telemetry for high-resolution high-cadence observations during a few hours of the window.

\subsection{Criterion 2: Objectives requiring Metis \& SoloHI to observe Earth-directed transients}

The objectives relative to the structure and propagation of CMEs and blobs ideally need Solar Orbiter and Earth in quadrature with SoloHI looking towards Earth, so Solar Orbiter at GSE -Y (i.e. the angle between Earth-Sun-Orbiter being of the order of 90$^\circ$. GSE is the Geocentric Solar Ecliptic system: it has its X axis towards the Sun and its Z axis perpendicular to the plane of the Earth's orbit around the Sun. This system has the advantage of being fixed with respect to the Earth-Sun line.) This criterion will preferably be applied at perihelia but also during high-latitude windows. Alternatively, it will be interesting to observe at other separation angles (45-90 degrees).
The SOOPs that are most suitable to run during these times are: \hyperlink{coronaldynamics}{L\_FULL\_HRES\_HCAD\_Coronal-Dynamics} (focused on the off-limb corona up to Earth) and \hyperlink{eruptionwatch}{L\_FULL\_HRES\_HCAD\_Eruption-Watch}. The second SOOP has higher telemetry demands but will be helpful if the CME happens to come towards Solar Orbiter: then it would be viewed sideways from Earth. RSWs that fall close to equinox would be preferred since Earth-directed CMEs (and southward IMF solar wind-magnetosphere coupling in general) tend to be more geoeffective at that time. Other SOOPs that will be run a few times at quadrature are  \hyperlink{cmeseps}{L\_FULL\_HRES\_MCAD\_CME-SEPs} and \hyperlink{flareseps}{L\_FULL\_MRES\_MCAD\_Flare-SEPs} (this one with SoloHI towards Earth). Even though these can be run at all times, some of the related sub-objectives benefit from quadrature with Earth, so that Earth-based observatories can observe the structure of the CME heading towards Solar Orbiter.

\subsection{Criterion 3: Slow solar wind connection science requiring Earth context for modelling prior to a remote-sensing window}
 
As already discussed, connection science objectives constitute a major goal of the Solar Orbiter mission. For planning purposes, we consider two different types of connection science campaigns: during the solar minimum and during the rest of the solar cycle.  

\paragraph*{During solar minimum.} The magnetic field configuration is much simpler than during the rest of the solar cycle, with slow solar wind coming from the streamer belt. In addition, during the early orbits of the mission, Solar Orbiter will stay close to the ecliptic. If SO/PHI observes the far side magnetic field in good resolution, and we combine that with the Earth-side magnetic field, the full solar magnetic field configuration can be modelled including the location of the Heliospheric Current Sheet that will determine the hemisphere Solar Orbiter will be connected to. This model will be the ideal starting point to run a longer term (10-20 days) connection SOOP using synoptic data of both in-situ and remote-sensing payload pointed to the most likely connection point. During this campaign, SO/PHI takes regular full disc magnetograms to update the magnetic field model as we go. The modelling will also improve as Earth and Solar Orbiter see overlapping longitude ranges of the Sun. The relevant SOOPs are \hyperlink{magnfieldconfig}{L\_FULL\_HRES\_LCAD\_MagnFieldConfig} for the magnetic field modelling (during the first remote-sensing window), while during the connection observations we will use \hyperlink{farsideconnection}{L\_BOTH\_MRES\_MCAD\_Farside-Connection}, possibly combined with \hyperlink{connectionmosaic}{L\_SMALL\_MRES\_MCAD\_Connection-Mosaic}.

\paragraph*{During the rest of the solar cycle.}  As the Sun becomes more active, the magnetic field modelling will become more challenging \citep{Rouillard2020a}. In these periods, we will employ more Earth observations to get a well-constrained model of the field that Solar Orbiter is going to fly through. If SO/PHI data are restricted or not available, we mainly rely on Earth to produce the model four days in advance due to VSTP turn-around loop. For this to happen, we need Solar Orbiter in the GSE sector X<1\,au and Y<0\,au, that is similar orbits than the ones needed for Earth-directed transients above. The further Solar Orbiter moves away from that sector, the more we rely on SO/PHI data to model the most likely connection point. Due to the more complicated and less reliable magnetic field modelling, we may want to use pointing mosaics to establish the most likely connection point using the SOOP \hyperlink{connectionmosaic}{L\_SMALL\_MRES\_MCAD\_Connection-Mosaic}. During later orbits, the concatenated perihelion and north windows will span a large range of latitudes over a short period of time. This SOOP will be combined with \hyperlink{slowwindconnection}{L\_SMALL\_HRES\_HCAD\_Slow-Wind-Connection}, involving high resolution and cadence remote-sensing observations to explore source regions in detail, or \hyperlink{ballisticconnection}{L\_SMALL\_MRES\_MCAD\_Ballistic-connection}. If we point at a coronal hole boundary, the SOOP \hyperlink{chboundaryexpansion}{L\_BOTH\_HRES\_LCAD\_CH-Boundary-Expansion} fits as well.

\subsection{Criterion 4: Polar objectives}

The different objectives that require high latitude (for example, all those that require observing the poles) have to be planned during high-latitude windows and split between objectives that need good telemetry and those that do not, to find the best suitable schedule. In addition, since solar global field reversal occurs during the maximum of activity, we will prioritise a relevant remote-sensing campaign at the maximum to characterise better the reversal process.

\subsection{Criterion 5: Opportunities for long-term remote-sensing observations}

Some science objectives benefit from a longer period of continuous remote-sensing observations, typically in synoptic mode. A special case of this criterion are remote-sensing observations from pole to pole with minimal interruption. These observations will be scheduled at times when two RSWs are concatenated or when there is minimal interruption between windows. However we need to note that helioseismology objectives require long, continuous observations at high temporal cadence, so that the synoptic mode would not be sufficient in that case.

\subsection{Criterion 6: Fast wind connection}

The SOOP \hyperlink{fastwind}{L\_SMALL\_HRES\_HCAD\_Fast-Wind} addresses two main science goals that in general need coronal holes as a target. The science objective regarding the sources of the fast solar wind will benefit from a low-latitude (or extended) coronal hole, to increase the chance of connection and to compare the composition of low and fast solar wind streams: this is most likely to happen in the declining phase of the solar cycle \citep{2015LRSP...12....4H}. We prefer orbits with a fast scan through a big range of latitudes (similar to the opportunities above for pole-to-pole observations --- criterion 5). In particular, to address the science goal regarding the origin of jets from polar coronal holes, high latitude windows are preferred to ensure the presence of a well-established polar coronal hole that can be observed in full. Limb pointing from medium latitude is also interesting to get the Doppler velocity component from SPICE combined with EUI for off-limb intensity. Observations from up close will be a big asset as well.

\subsection{Criterion 7: Science objectives needing perihelia but lower telemetry requirements}

The SOOP \hyperlink{flareseps}{L\_FULL\_MRES\_MCAD\_Flare-SEPs} needs medium telemetry downlink. Some of its sub-objectives require quadrature with Earth (i.e. the angle Earth-Sun-Orbiter being in the order of 90$^\circ$), so this SOOP is also mentioned above in Criterion 2. The SOOP \hyperlink{lisstix}{L\_IS\_STIX} needs low telemetry (in practice this SOOP is likely to run throughout all RSWs). As the related telemetry needs of these SOOPs are moderate to low, we will schedule them during outbound perihelia, that is those where Solar Orbiter is moving away from Earth so downlink performance is decreasing.

\subsection{Criterion 8: Global magnetic field reconstruction}

RSWs at the far side of the Sun will be used to have regular, low cadence imaging of magnetic field, to allow global field reconstruction. This goal will be addressed by the SOOP \hyperlink{coronalsynoptic}{L\_FULL\_LRES\_MCAD\_Coronal-Synoptic} or \hyperlink{magnfieldconfig}{L\_FULL\_HRES\_LCAD\_MagnFieldConfig}. Ideally, we plan these SOOPs at regular far-side windows covering a wide range of phases in the solar cycle. In addition, the same opportunities will allow \hyperlink{coronalsynoptic}{L\_FULL\_LRES\_MCAD\_Coronal-Synoptic} to address the study of the magnetic field at high-latitude spanning different longitudes.

\subsection{Criterion 9: Remaining objectives and special circumstances}

Several objectives and their related SOOPs do not fit in the above criteria and must be planned separately. Some examples can be found in the following paragraphs.

\paragraph*{Abundance of minor ions as a function of height in the corona as an indicator of slow or fast wind.} This objective will be addressed through the SOOP \hyperlink{compositionvsheight}{R\_SMALL\_HRES\_LCAD\_Composition-vs-Height}. For this goal, we need limb pointing of either an active region (with open field at the edges) or the boundary of a streamer, so the target will be chosen at the time of VSTP. A perihelion is preferred but not required. Running this SOOP at a slightly larger distance from the Sun will benefit from Metis participation and a wider field of view for SPICE. Exactly the same requirements are needed for other objectives, such as the study of the role of shocks in generating SEPs, which also needs limb pointing during a remote-sensing window outside 0.55\,au, so that Metis will contribute to the observations.

\paragraph*{Resolve the geometry of fine elemental loop strands.} This objective will be addressed by the SOOP \hyperlink{finescalestructure}{R\_SMALL\_HRES\_LCAD\_Fine-Scale-Structure}. This needs the highest possible resolution at close perihelia, but no high cadence and thus no particularly high telemetry needs. All close perihelia within 0.3\,au seem to be good opportunities for this SOOP.

\paragraph*{Study of density fluctuations in the extended corona as a function of the outflow velocity of the solar wind.} This objective will be addressed by the SOOP \hyperlink{densityfluctuations}{R\_FULL\_HRES\_HCAD\_Density-Fluctuations}. Metis and SoloHI drive this SOOP, which needs to be repeated at several distances, that is at each remote-sensing window, but not too far out for Metis to still see the density fluctuations. Eight hours per window are sufficient.

\paragraph*{Photospheric dynamics.} This objective will be addressed by the SOOP \hyperlink{atmosphericdynamicsstructure}{R\_SMALL\_HRES\_HCAD\_Atmospheric-Dynamics-Structure}, which involves high telemetry needs for EUI, SO/PHI and SPICE during a short time (up to one hour). We need either perihelion for quiet Sun or close-in high-latitude windows for coronal holes (i.e. north windows). For perihelion windows, we will select the same ones as for the remote-sensing burst above.

\paragraph*{Active Region dynamics.}  The best opportunities to study CME initiation and structure (close to the Sun), are to point to active regions at perihelion with the SOOP \hyperlink{ardynamics}{R\_SMALL\_HRES\_HCAD\_AR-Dynamics}. We prefer coordinated observations with ground-based or space-borne instruments at 1\,au for modelling and obtaining CME context.

\paragraph*{Spatial distribution of SEPs in the inner heliosphere.} This will be addressed by the SOOP \hyperlink{flareseps}{L\_BOTH\_MRES\_MCAD\_Flare-SEPs}. It needs many events, ideally observed from different viewpoints (including Earth and other viewpoints) and different distances (e.g. from Parker Solar Probe or STEREO). It also needs a range of latitudes (some high-latitude windows as well). Science closure of the relevant objectives requires a large number of time intervals using this SOOP.

\paragraph*{Energy flux in the lower atmosphere.} In order to better understand coronal heating we will benefit from coordinated observations with Earth-based (DKIST) and near-Earth (IRIS) facilities, with sets of particular geometries between Solar Orbiter, the target on the Sun, and Earth.

\paragraph*{Helioseismology.} This technique requires very long (from days to months) continuous series of Dopplergrams at one minute cadence from PHI FDT and HRT. The spatial resolution will be adapted depending on the science objectives, see SOOPs \hyperlink{globalhelioseismology}{R\_FULL\_LRES\_HCAD\_Full-Disc-Helioseismology} and \hyperlink{localareahelioseismology}{R\_FULL\_LRES\_HCAD\_Local-Area-Helioseismology}.

\paragraph*{Objectives that could be enhanced with observations from the Parker Solar Probe.} Whenever possible coordinated observations with Parker Solar Probe will be planned \citep{Velli2020a}. Examples of SOOPs that would benefit from such observations are \hyperlink{flareseps}{L\_BOTH\_MRES\_MCAD\_Flare-SEPs}, \hyperlink{lissolohistix}{L\_IS\_SoloHI\_STIX} and \hyperlink{probequadrature}{L\_FULL\_LRES\_MCAD\_Probe-Quadrature}, which requires Parker Solar Probe in quadrature with Solar orbiter (i.e. the angle Probe-Sun-Orbiter being of the order of 90$^\circ$).

\section{Remote-sensing observations outside of the remote-sensing windows}

The nominal strategy of RSWs centred on the perihelia and maximum/minimum latitudes, as described in Section 3, maximises the scientific return of the mission during these unique intervals, but also implies that the connection between the remote-sensing and the in-situ observations is effective only during about one-sixth of the mission duration. In order to increase the duration of the joint observations, the instruments and operations teams have explored the possibility for the remote-sensing instruments to dedicate a fraction of their telemetry allocation to perform synoptic `out-of-window' observations that will provide the contextual information necessary to enable connection science throughout orbits, while maintaining a low-resource and low-impact profile and avoiding violating the Electro-Magnetic Cleanliness (EMC) requirements of the in-situ instruments. While added late to the development of the mission ground segment, out-of-window observations can be made almost resource neutral for the ground operations teams as long as they fit into the low-latency data volume (see \cite{Auchere2020a} and \cite{Sanchez2020}). Each remote-sensing instrument therefore designed a `synoptic' type programme that can be run continuously to provide basic contextual information without impacting their core objectives. This includes a highly-compressed 15-min cadence full disc image from EUI/FSI, daily full Sun line-of-sight magnetic field and continuum images from SO/PHI, a 30-min cadence visible light image from Metis, a 30-min cadence 2.5\textdegree-wide equatorial and latitudinal swaths from SoloHI, one daily first-ionisation-potential (FIP) bias map from SPICE and nominal observations from STIX. Even though this programme is not yet approved by ESA, it is currently under study for an implementation during the nominal mission phase starting in November 2021.

\section{An example of Long Term Planning}

In this section we describe an example of a 6-month period for which we have applied the above strategy to produce the Long Term Plan. This example covers the period between July 1, 2025 and December 31, 2025. Although this particular period is considered only for illustrative purposes, it is selected as an example of the fact that the spacecraft trajectory is suitable for the scheduling of many different SOOPs and how to apply the strategy in cases for which an informed choice is needed about the science objectives to be addressed. Furthermore this period was exercised within the Solar Orbiter instruments teams and the Science Operations Centre before launch, producing a detailed observation plan and resource simulations. The basic results are described in this section, but we note that this is just an example and that the actual plan for these dates will change in the next years following discussions within the Science Working Team. 

\begin{figure}
\includegraphics[width=\columnwidth]{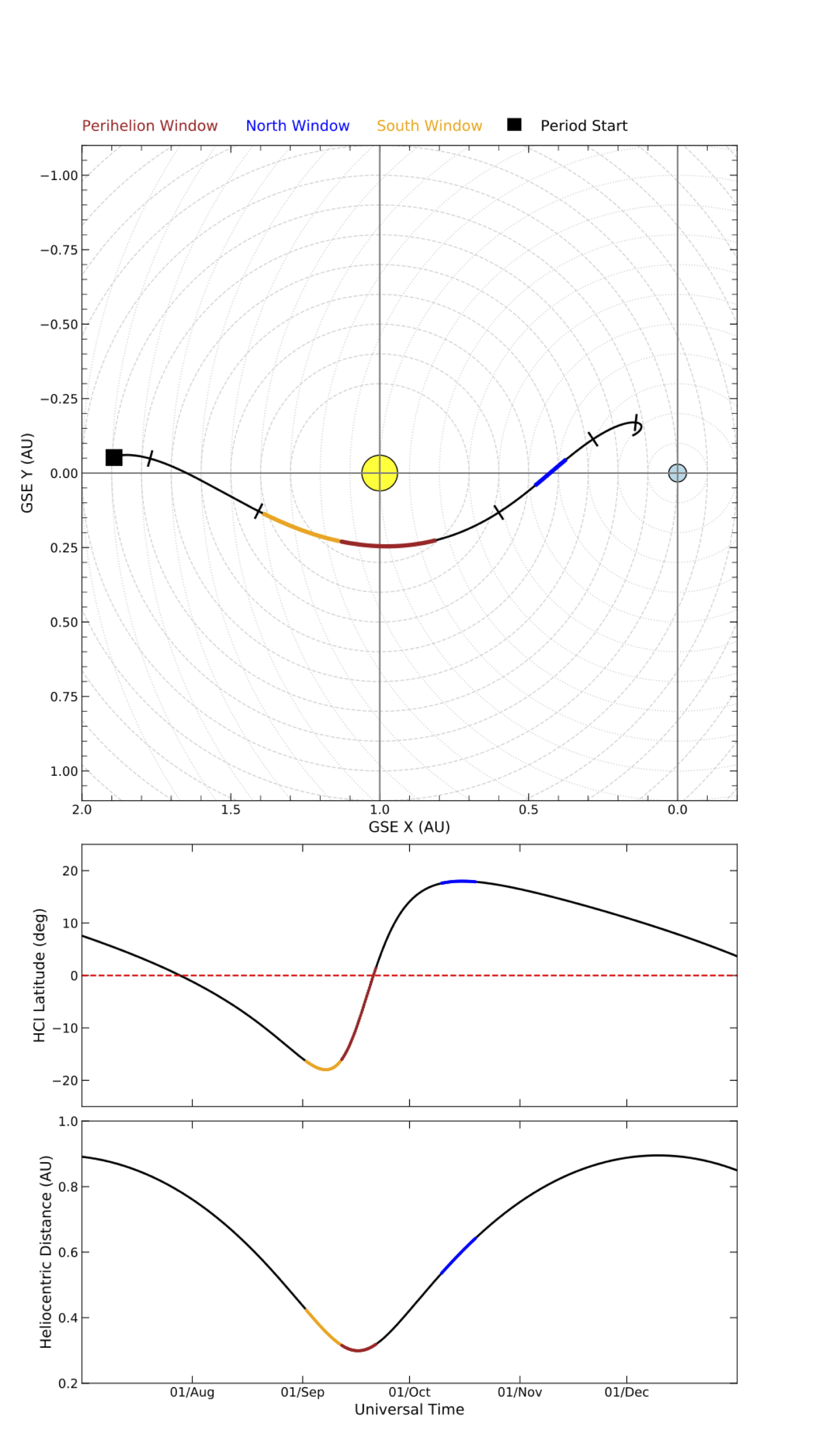}
\caption[]{Top panel: trajectory of the spacecraft between July 1, 2025 and December 31, 2025 projected onto the GSE XY plane (the Sun is depicted by the yellow circle at coordinates X=1\,au and Y=0\,au, which is the distance from Earth placed at the origin of the system and depicted by the pale blue circle). The first day of the trajectory corresponds to the black square on the left of the plot. The two bottom panels show the heliographic latitude and the heliocentric distance. The sections of the curves corresponding to three plausible remote-sensing windows are coloured in yellow, red and blue. Black corresponds to periods during which only in-situ instruments are operating without remote-sensing observations.}
\label{LTP-example}
\end{figure}

On the top panel of Fig.~\ref{LTP-example}, we see the trajectory of the spacecraft between July 1, 2025 and December 31, 2025 projected onto the GSE XY plane (the Sun is depicted by the yellow circle at coordinates X=1\,au and Y=0\,au, which is the distance from Earth placed at the origin of the system and depicted by the pale blue circle). In this system, Earth is fixed with respect to the Sun. The heliographic latitude and the heliocentric distance are shown as a function of time on the two bottom panels. The first day of the trajectory corresponds to the black square on the left of the plot. The spacecraft reaches a latitude of almost 20 degrees and the perihelion is as close as 0.3\,au. 
The first decision that needs to be taken for every such period is where to place the three 10-days RSWs (in this example the sections of the curves corresponding to the three different windows are coloured in yellow, red and blue). The sections of the curves that are black correspond to periods during which only in-situ instruments are operating without remote-sensing observations. As a default baseline, the choice is to centre one 10-day window at the perihelion (red part) and the other two windows at the highest north (blue) and south (yellow) latitudes. Depending on the science goals that we need to address, a slightly different configuration can be adopted, for example, by concatenating two of the windows around the perihelion, or even all three windows together if we need to perform continuous remote-sensing observations for 30 consecutive days. In this case, the two first windows have been concatenated, but the configuration is such that they also preserve their original characteristics (the first window is still at high latitude starting with three days of delay compared to its default placement and the second one is centred around perihelion). The first window (RSW1) starts on September 5, 2025 and ends on September 14. The second window (RSW2) starts on September 15 and ends on September 24. The third window (RSW3) spans the period between 9 and 19 of October. There is in principle margin to place RSW3 in a different part of the orbit (earlier or later), but it is decided to leave it at the highest latitude, which is the baseline operational scenario. This is particularly important for this specific period, since it is only the second period that the spacecraft would be at such a high latitude. 

Before examination of the criteria described in Section 5, we need to thoroughly consider what has already been planned and executed in the previous planning periods. Since this is an example, we just proceed with the following overview of the previous periods and operational assumptions. Prior to this period we will have had seven other planning periods with perihelia around 0.3\,au. The previous period is the first one when the spacecraft would have reached a latitude higher than 15 degrees. All previous RSWs at perihelia will have encompassed quadratures between Solar Orbiter and Earth and there will have been six previous RSWs where at least 90 degrees in longitude of the far side of the Sun will have been visible to Solar Orbiter.

Examination of the first criterion of the strategy described in the previous section (high-cadence remote-sensing observations) requires an assumption on the mass memory fill state at the beginning of the period as well as the telemetry resources. Given at  the start of the period Solar orbiter is on the far side of the Sun from Earth (superior conjunction), it is reasonable to assume that the mass memory is half full since it would contain data from the previous period that did not have an opportunity to be downlinked. With this assumption, it appears difficult to perform high-resolution and high-cadence observations, so that the first criterion leads us to the conclusion that it is preferable to avoid planning the remote-sensing burst SOOP or only plan it for a few days, if needed.

Regarding the second criterion, since the spacecraft mostly lies in GSE +Y, SoloHI would not be in a position to observe Earth-directed transients. However we can use Metis and ground- or L1-based assets to observe CMEs directed towards the spacecraft at perihelion. It is therefore reasonable to run the SOOP \hyperlink{eruptionwatch}{L\_FULL\_HRES\_HCAD\_Eruption-Watch} during RSW2 and/or L\_FULL\_HRES\_MCAD\_CME-SEPs and \hyperlink{flareseps}{L\_BOTH\_MRES\_MCAD\_Flare-SEPs}. 

For the criteria 3 and 6, which concern the slow and fast wind connection, we need to take into account that this period corresponds to the maximum of solar activity and, therefore, it will not be easy to trace the sources of the slow wind. We could however plan \hyperlink{fastwind}{L\_SMALL\_HRES\_HCAD\_Fast-Wind} for RSW3. We could also try to trace the slow wind with the SOOP \hyperlink{slowwindconnection}{L\_SMALL\_HRES\_HCAD\_Slow-Wind-Connection} during RSW2 and RSW3 for wind coming from coronal holes and active region boundaries or run the synoptic campaign \hyperlink{poletopole}{L\_BOTH\_LRES\_MCAD\_Pole-to-Pole} during RSW1 and RSW2. 

The concatenation of the two first windows enables us to consider the implementation of criterion 5 for long-term remote-sensing observations, including the fact that in this particular example we can have a pole-to-pole coverage, running the SOOP \hyperlink{poletopole}{L\_BOTH\_LRES\_MCAD\_Pole-to-Pole} for RSW1 and RSW2. 

In summary, during this period we address polar science objectives during the first five days of RSW1 at high latitude and high resolution together with the possibility of off-pointing. During the following 15 days (second half of RSW1 and RSW2), we observe the full solar disc from one pole to another together with the possibility of catching an eruption as well as high-resolution observations with the remote-sensing instruments as a secondary goal. We end this period with the third window running ten days of solar wind tracing and possibly polar science at the other pole. 
Criterion 8 regarding the global magnetic field reconstruction could also be applicable during RSW1 for observations of the far-side of the Sun. Moreover, we note that not all criteria are considered in this example, since this depends on the overall planning of the entire nominal mission phase. For instance, if objectives of polar science have been tackled during the previous planning period, then we might not consider it for this one. It is important to note that this limitation is only inherent to the restricted nature of this example, which is described here as an isolated period and not part of the entire mission, as it should be. When the entire plan is considered, the Science Working Team makes sure that all objectives and all SOOPs are adequately placed for a sufficient number of times throughout the nominal and the extended mission phases. In practice, after the Science Working Team builds the whole plan, coordinators for each SOOP and for each individual period will be nominated and have the responsibility to work out the details of the science to be addressed together with all instrument teams and interested scientists as well as refining the operational aspects in coordination with the Science Operations Centre and the Project Scientists.

\section{Cruise phase}
The above operational concept and planning strategy are only valid after the nominal mission phase starts in November 2021, when the spacecraft perihelion gets close enough to make novel remote-sensing observations. During the cruise phase (planned between June 2020 and November 2021), only the in-situ instruments are operating on a best-effort basis and with limited telemetry. The remote-sensing instruments are off, except for specific periods called ``remote-sensing checkout windows'' (RSCWs). These windows are different from RSWs discussed so far because they are not intended for scientific observations. They are instead focused on the calibration and characterisation of the remote-sensing instruments and on all observing activities needed to prepare those instruments to be fully operational by November 2021. These checkout windows had to be planned taking into account mission and platform restrictions, calibration opportunities and instrument specific limitations. The current planning consists of four checkout windows on 17-22 June 2020, 20-25 February 2021, 21-24 March 2021 and 22-30 September 2021 covering different distances and therefore different thermal environments. The planning for the in-situ instruments is rather straightforward throughout the cruise phase (normal mode with scheduled bursts) and they are operating continuously as during the nominal and extended mission phases. Special coordination is ongoing with other missions (e.g. Parker Solar Probe and BepiColombo) to take advantage of unique opportunities for joint observations (see \citealt{Velli2020a}).  

\section{Summary}

Given the nature and complexity of the Solar Orbiter mission, it is essential to implement a plan for the entire mission and an operational strategy to make sure that all science goals will be tackled in the most optimal way when the opportunities arise. Most of Solar Orbiter's science objectives are challenging not only from the science point of view, but also from a technical perspective, since many objectives demand high-resolution and high-cadence remote-sensing observations that have to be accommodated within the limited spacecraft resources. To achieve this, many members of the international solar physics and heliophysics community have contributed to the generation of the Science Activity Plan covering all mission phases throughout the next decade. This plan makes extensive use of the Solar Orbiter Observing Plans (SOOPs), which are the building blocks of instrument modes required to address multiple science goals. Inversely, the SAP will be used by the community as a tool to work through detailed science activities that will lead us to the answers to the top-level mission objectives. It is important to stress that this is a dynamic and ever-evolving plan as new data and scientific knowledge become available. All decisions regarding the planning will be reviewed and re-considered regularly. For this reason the participation and input of all scientists are welcome to ensure the maximum scientific return from Solar Orbiter and to advance significantly humankind's understanding of our star.

\begin{acknowledgements}
Solar Orbiter is a space mission of international collaboration between ESA and NASA with contributions from national agencies of ESA member states. The spacecraft has been developed by Airbus and is being operated by ESA from the European Space Operations Centre (ESOC) in Darmstadt, Germany. Science operations are carried out at ESA's European Space Astronomy Centre (ESAC) in Villafranca del Castillo, Spain. SWA is an international collaboration which has been funded by the UKSA,
CNES, ASI, NASA and the Czech contribution to the ESA PRODEX programme. UAH authors want to thanks the Spanish MINECO-FPI-2016 predoctoral grant with FSE, and its project FEDER/MCIU-AEEI/Proyecto ESP2017-88436-R. The Spanish contribution to SO/PHI has been funded by the Spanish Ministry of Science and Innovation through several projects, the last one being RTI2018-096886-B-C5, and by ``Centro de Excelencia Severo Ochoa'' programme under grant SEV-2017-0709. RAH, RCC, DMM, SPP,  and AV acknowledge the support of the NASA Heliophysics Division, Solar Orbiter Collaboration Office under IAT NNG09EK11I. JEP acknowledges grant UKRI/STFC ST/N000692/1. All French involvements are supported by CNES and CNRS. DV is supported by the STFC Ernest Rutherford Fellowship ST/P003826/1 and STFC Consolidated Grant ST/S000240/1.
The authors thank the referee for her constructive comments and suggestions, which led to the substantial improvement of this paper.
\end{acknowledgements}	


\bibliographystyle{aa}
\bibliography{references}

\end{document}